\documentclass[aps,pre,reprint,groupedaddress]{revtex4-1}

\usepackage{amsmath}
\usepackage{amssymb}
\usepackage{graphicx}
\usepackage{siunitx}
\usepackage{xcolor}

\newcommand{\dif}{\mathrm{d}}
\newcommand{\dd}[2]{\frac{\dif #1}{\dif #2}}
\newcommand{\bigO}{\mathrm O}

\newcommand{\mf}{\mathrm{mf}}

\DeclareMathOperator{\qpoly}{\mathcal H}
\DeclareMathOperator{\qint}{\mathcal Q}
\DeclareMathOperator{\erf}{erf}

\newcommand\pp[2]{\frac{\partial #1}{\partial #2}}
\newcommand{\ave}[1]{\left< #1 \right>}

\renewcommand{\vec}[1]{\boldsymbol #1}
\newcommand{\mat}[1]{\boldsymbol{\mathsf #1}}

\newcommand{\drop}[1]{}

\newcommand{\comment}[1]{{#1}}

\begin{document}


\title{A Model of Electron Transport in Dense Plasmas Spanning Temperature Regimes}


\author{Nathaniel R. Shaffer}
\email[]{nshaffer@lanl.gov}
\affiliation{Los Alamos National Laboratory}

\author{Charles E. Starrett}
\affiliation{Los Alamos National Laboratory}


\date{\today}

\begin{abstract}
  We present a new model of electron transport in warm and hot dense plasmas which combines the quantum Landau-Fokker-Planck equation with the concept of mean-force scattering.
  We obtain electrical and thermal conductivities across several orders of magnitude in temperature, from warm dense matter conditions to hot, nondegenerate plasma conditions, including the challenging crossover regime between the two.
  The small-angle approximation characteristic of Fokker-Planck collision theories is mitigated to good effect by the construction of accurate effective Coulomb logarithms based on mean-force scattering, which allows the theory to remain accurate even at low temperatures, as compared with high-fidelity quantum simulation results.
  \comment{Electron-electron collisions are treated on equal footing as electron-ion collisions.
  Their accurate treatment is found to be essential for hydrogen, and is expected to be important to other low-$Z$ elements.}
  We find that electron-electron scattering remains influential to the value of the thermal conductivity down to temperatures somewhat below the Fermi energy.
  The accuracy of the theory seems to falter only for the behavior of the thermal conductivity at very low temperatures due to a subtle interplay between the Pauli exclusion principle and the small-angle approximation as they pertain to electron-electron scattering.
  Even there, the model is in fair agreement with \textit{ab initio} simulations.
\end{abstract}


\maketitle

\section{Introduction}
\label{sec:intro}

Accurate prediction of the electrical and thermal conductivity of dense plasmas is an ongoing challenge, with a decades-long history mainly in the fields of stellar modeling\cite{HubbardApJS1969,FlowersApJ1976,CassisiApJ2007} and inertial confinement fusion (ICF)\cite{AtzeniMeyerTerVehn,HuPRE2014b}.
In direct-drive ICF, thermal conduction by the electrons is the main means by which laser energy is transferred from the ablated plasma to the fuel capsule.
The ICF fuel hot spot can also lose energy due to thermal conduction with the colder surrounding fuel.
In stellar evolution models, the electron thermal conductivity is important to the evolution of low-mass stars and the cooling of white dwarfs.
Dense plasmas typically have densities ranging from a fraction of solid density to several times solid density, with temperatures from a few \SI{}{\electronvolt} to a few \SI{}{\kilo\electronvolt}.
The challenge in modeling conduction at these conditions comes in two forms.
First, one needs a kinetic theory that is able to account for strong Coulomb interactions between the ions, which may be partially ionized, as well as Fermi degeneracy effects in the electrons.
Second, one needs realistic collision cross-sections that not only reflect these influences, but also account for the internal electronic structure of the ions, which is strongly temperature dependent at dense plasma conditions.

The extreme cases of low temperature and high temperature are reasonably well-understood, and the conductivities are given by the Ziman theory of liquid metals\cite{ZimanPM1961}, and the Spitzer-H\"arm theory of classical plasmas\cite{SpitzerPR1953}, respectively.
In between these two extremes is several orders of magnitude in temperature over which neither approach is well-justified.
The Ziman approach includes strong-coupling effects in the ions and Fermi statistics for the electrons, but neglects electron-electron scattering entirely, which is important in low-$Z$ materials at high temperatures.
The Spitzer-H\"arm approach includes the influence of electron-electron scattering on equal footing with electron-ion scattering, but it is only valid for hot plasmas where the ions are weakly coupled and the electrons are nondegenerate.

The need for new theoretical predictions of dense plasma conductivity is underscored by the fact that quantum simulations of electron transport in plasmas become impractical and possibly unreliable at high temperatures.
The prevailing methodology, Kohn-Sham molecular dynamics paired with the Kubo-Greenwood approximation (``QMD''), scales prohibitively with increasing temperature\cite{SjostromPRL2014}.
A recent QMD study on hot dense hydrogen by Desjarlais et al.\cite{DesjarlaisPRE2017} found that it is not only extremely difficult to achieve numerical convergence of the transport coefficients at high temperatures but that the resulting predictions for the thermal conductivity are systematically too large due to an incomplete account of electron-electron scattering.
The precise nature of this error is an open question, but it is closely related to the Kubo-Greenwood approximation\cite{ReinholzPRE2015,DuftyCPP2018,DesjarlaisPRE2017}.
This electron-electron scattering error may be resolved by methods which go beyond the Kubo-Greenwood approximation such as time-dependent density functional theory\cite{BaczewskiPRL2016,AndradeEPJB2018} or $GW$ corrections\cite{HollebonPRE2019,FaleevPRB2006}, but these methods are also not yet practical at high temperatures (although we note recent advances within orbital-free\cite{SjostromPRL2014,SjostromPRE2015,DingPRL2018,WhitePRB2018} and stochastic density functional theory\cite{CytterPRB2018} approaches).
This leaves kinetic theory as the only practical avenue for investigating electron transport in dense plasmas across temperature regimes from degenerate to classical.

In this paper, we present a new electron transport model that combines the quantum Landau-Fokker-Planck (qLFP) kinetic theory with the concept of mean-force scattering.
The qLFP theory is a generalization of the classical Fokker-Planck equation used by Spitzer and H\"arm that accounts for quantum statistics, which is necessary to capture the effect of Pauli blocking at high densities\cite{DanielewiczPA1980,DaligaultPOP2016}.
Like its classical counterpart, the qLFP theory is formally limited to weakly coupled plasmas, where transport happens mainly via glancing collisions.
We extend its domain of accuracy to lower temperatures by constructing Coulomb logarithms based on the concept of mean-force scattering, where the scattering cross-sections are calculated using the potential of mean force as the interaction potential\cite{BaalrudPRL2013,DaligaultPRL2016,ShafferPRE2017,StarrettHEDP2017,BaalrudPOP2019}.
\comment{At high temperatures, the potential of mean force reduces to the Debye-H\"uckel potential (static mean-field screening)\cite{HansenMacDonald, BaalrudPRL2013}.}
At low temperatures, the potential of mean force models how screening and inter-particle correlations influence the effective pairwise interaction.
This allows for the account of strong coupling effects in a plasma kinetic framework, which enables the theory to be used at both low and high temperature.

The combined account of strong coupling effects, arbitrary electron degeneracy, and electron-electron scattering is the main strength of our approach compared with other recently developed quantum kinetic transport models, which so far include only two of the three effects.
\comment{We mention here only a few of the most recent models; for comprehensive bibliographies on earlier dense plasma conductivity literature, we refer the reader to Refs.~\cite{StygarPRE2002} and~\cite{RedmerPR1997}, as well as a recent comparative study on the AC conductivity by Veysman et al.~\cite{VeysmanPRE2016}.}
The semiclassical Lenard-Balescu model by Whitley et al.\cite{WhitleyCPP2015} is intended for hot dense plasmas.
They account for the wavelike nature of electrons as well as electron-electron scattering but do not include Pauli blocking or correlation effects which are important at low temperatures.
\comment{Reinholz et al.\cite{ReinholzPRE2015} derived an electron-electron scattering correction to the electrical conductivity of a Lorentz plasma, but that work does not consider the thermal conductivity.
Their approach is couched in the Zubarev linear response formalism\cite{RopkePRA1988,RopkeBook}, which in principle accounts for ion correlations, electron-electron collisions, and Pauli blocking in addition to other effects not considered here like electron-atom collisions~\cite{AdamsPRE2007,RosmejPRE2017}.
In their practical results, Reinholz et al.~adopted a dynamically screened Born approximation similar to Whitley et al.'s Lenard-Balescu approach, except that their calculations account for Pauli blocking.
However, neither model's final results account for correlation and screening effects beyond the dynamic random phase approximation.}
Both models also neglect the exchange channel for electron-electron scattering, which is unimportant at high temperatures but can become significant for $T \lesssim 10T_F$\cite{DaligaultPRL2017}.
The relaxation-time model by Starrett\cite{StarrettHEDP2017} includes Pauli blocking and accounts for correlations using a mean-force scattering cross-section for electron-ion collisions, but electron-electron collisions are accounted for only through the correction formula proposed in Ref.~\cite{ReinholzPRE2015}.
Starrett also does not consider the thermal conductivity.

The remainder of the paper is organized as follows.
Section~\ref{sec:theory} reviews some essential aspects of the qLFP kinetic theory, the approximations involved in using it to predict transport coefficients, as well as our model for extending its range of applicability using accurate Coulomb logarithms.
Section~\ref{sec:results} contains our results for the electrical and thermal conductivity, emphasizing the temperature dependence in compressed hydrogen plasma and solid-density aluminum plasma.
Comparisons with other models and available QMD data are made.
Section~\ref{sec:discussion} contains detailed discussion of some important features of our results, including the occurrence of a minimum conductivity, the role of electron-electron scattering, and a subtle consequence of the small-angle approximation in the qLFP theory.
Section~\ref{sec:conc} offers some concluding remarks and directions for continued investigation.

\section{Theory}
\label{sec:theory}

\subsection{The Quantum Landau-Fokker-Planck Theory}
\label{sec:qlfp}

The qLFP kinetic equation governs the evolution of the one-particle phase-space distribution functions, $f_i$, for a mixture of $K$ species.
It reads\cite{DanielewiczPA1980,DaligaultPOP2016}
\begin{equation}
  \label{eq:kin-eq}
  \left[\pp{}{t} + \frac{\vec p}{m_i}\cdot\pp{}{\vec r} + \vec F_i \cdot\pp{}{\vec p}\right] f_i(\vec r, \vec p, t)
  = \sum_{j=1}^K C_{ij}
\end{equation}
where $\vec F_i$ is the external force acting on species $i$ (we consider only static, uniform electric fields here), and $C_{ij}$ is the collision operator
\begin{multline}
  \label{eq:collop}
  C_{ij}[f_i, f_j] = \Gamma_{ij} \pp{}{\vec p}\cdot\int \mat K \cdot \left[
    \pp{f_i}{\vec p} f_j(\vec p') \bar f_j(\vec p')
  \right. \\ - \left.
    \pp{f_j}{\vec p'} f_i(\vec p) \bar f_i(\vec p) 
  \right]
  \frac{d\vec p'}{\omega_j}
\end{multline}
in which $\bar f_i(\vec p) = 1 + \eta_i f_i(\vec p)$, with $\eta_i = -1,0,1$ for fermions, classical particles, or bosons, respectively.
The prefactor
\begin{equation}
  \label{eq:gamma-factor}
  \Gamma_{ij} = 4\pi e_i^2 e_j^2 m_{ij} \ln\Lambda_{ij}
\end{equation}
involves the charge $e_i$ of each species, the reduced mass $m_{ij} = (m_i^{-1} + m_j^{-1})^{-1}$, and a Coulomb logarithm.
The tensor
\begin{equation}
  \label{eq:vmat}
  \mat K(\vec k) = \frac{1}{2\hbar k} \left( \mat I - \frac{\vec k \vec k}{k^2} \right)
\end{equation}
projects onto the plane in momentum space normal to the relative momentum, $\hbar\vec k = \vec p - \vec p'$.
Finally, $\omega_i = (2\pi\hbar)^3/g_i$ is the phase-space density per state, in which $g_i$ is the multiplicity of each state, i.e., 2 for electrons.
The distribution functions are normalized so that
\begin{equation}
  \label{eq:f-norm}
  n_i(\vec r, t) = \int f_i(\vec r, \vec p, t) \frac{d\vec p}{\omega_i}
\end{equation}
is the local density.
In local thermodynamic equilibrium, they are
\begin{equation}
  \label{eq:f-eq}
  f_i^{(0)}(\vec r, \vec p, t) = \left\{ e^{-\beta(\vec r, t)\left[\frac{p^2}{2m_i} - \mu_i(\vec r, t)\right]} - \eta_i \right\}^{-1}
\end{equation}
i.e., the Bose-Einstein, Maxwell-Boltzmann, or Fermi-Dirac distribution with inverse temperature $\beta=(k_BT)^{-1}$ and chemical potential $\mu_i$.

The approximations involved in using the qLFP collision operator can be understood from its relation to Uehling and Uhlenbeck's (UU) Boltzmann-like collision operator\cite{UehlingPR1933}
\begin{multline}
  \label{eq:uu}
  C^{UU}_{ij}[f_i,f_j] = \int [f_i(\hat{\vec p}) f_j(\hat{\vec p}') \bar f_i(\vec p) \bar f_j(\vec p')
  \\
  - f_i(\vec p) f_j(\vec p') \bar f_i(\hat{\vec p}) \bar f_j(\hat{\vec p}') ] v_{ij} d\sigma_{ij} \frac{d\vec p'}{\omega_j}
\end{multline}
where hats denote post-collision momenta, $d\sigma_{ij}$ is the differential cross-section\comment{, and $v_{ij}$ is the relative velocity.}
The qLFP collision operator follows by considering only collisions with small momentum transfer\cite{DanielewiczPA1980}.
As such, it inherits all the approximations of the UU theory (e.g, isolated binary collisions) in addition to the small-momentum-transfer approximation.

Formally, the distribution functions involved in the qLFP and UU theories are the one-particle Wigner distributions.
In the first equation of the Wigner hierarchy (Irving-Zwanzig equation), the external force term and the pair collision term involve non-local operators in space\cite{IrvingJCP1951}.
The derivation of the UU equation (and thus qLFP) from the Wigner hierarchy requires expanding these operators in powers of $\hbar$\cite{MoriPTP1952,RossJCP1954}.
Mori and Ono interpret this as a semi-classical treatment of diffraction, that the wavelength of electrons is assumed small compared to spatial variations in the external potential as well as compared to pair interaction length scales\cite{MoriPTP1952}.
In considering only weak external forces, the former is of no consequence, but the latter implies that the qLFP equation involves a semi-classical view of scattering inherited from the UU equation.
Another possible interpretation is that a binary collision picture requires the wavelength of electrons must be small compared to the effective range of interactions, otherwise the electrons will simultaneously diffract off many scatterers, the so-called multiple-scattering effect\cite{WilsonHEDP2011,StarrettPRE2018}.

  
\comment{In traditional Fokker-Planck theories}, Coulomb logarithms appear in the collision operator due to an assumption that collisions may be treated using weak Coulomb (or linearly screened) scattering~\cite{MontgomeryTidman,DanielewiczPA1980, DaligaultPOP2016}.
The logarithm results from cutoffs imposed to regularize a divergent integral over scattering angles (alternatively, momentum transfer).
The assumed $k^{-4}$ momentum dependence of the cross-section is a useful analytic simplification in the Chapman-Enskog solution of the qLFP equation\cite{DaligaultPOP2018}, but it is essentially a weak-scattering approximation that will break down at low temperatures and high densities if left uncorrected.
\comment{In the present work, we replace the usual Coulomb logarithms with new numerical values constructed from the cross-sections for mean-force scattering, which accounts for static screening in the plasma and removes the need to impose hard cutoffs on the momentum transfer.}
Such corrections are described in Section~\ref{sec:coul-log}.

\subsection{Coulomb Logarithms}
\label{sec:coul-log}

The values of the transport coefficients predicted by the qLFP theory depend on the model adopted for the Coulomb logarithms.
In textbook theory, Coulomb logarithms appear in the collision operator due to the application of the Rutherford scattering cross-section formula\cite{MontgomeryTidman}
\begin{equation}
  \label{eq:dcs-approx}
  d\sigma_{ij} \sim \left| \frac{e_i e_j m_{ij}}{\hbar^2 k^2 (1 - \cos\theta)} \right|^2 d(\cos\theta) d\phi
\end{equation}
whose associated transport cross-sections
\begin{align}
  \sigma^{(r)}_{ij}(k)
  &= \int (1 - \cos^r\theta) d\sigma_{ij}
    \label{eq:sig-r} \\
  &\sim 4\pi r \left| \frac{e_ie_jm_{ij}}{\hbar^2k^2} \right|^2 \int_0^{2k} \frac{dq}{q}
    \label{eq:ruth-transport}
\end{align}
diverge logarithmically at small momentum transfer $q^2=2k^2(1-\cos\theta)$.
Divergences of this sort are usually ameliorated by applying \textit{ad hoc} cutoffs.
There is an extensive literature arguing the plausibility of various cutoffs in hopes to recover some of the physics of diffraction, screening, and/or large-angle scattering that is left out when assuming Coulomb-like scattering.
Among the most widely used is the prescription of Lee and More\cite{LeePF1984}, who suggest taking
\begin{equation}
  \label{eq:lnL-LM}
  \ln\Lambda_{\mathrm{LM}} = \max\left\{2, \frac12 \ln\left(1 + \frac{b^2_{\max}}{b^2_{\min}}\right) \right\}
\end{equation}
with $b_{\min} = \max\{ \frac{Ze^2}{3k_BT}, \frac{\hbar}{\sqrt{12m_ek_BT}} \}$ and $b_{\max} = \max\{\lambda_D, a_I \}$
where $\lambda_D$ is the (total) Debye length and $a_I$ is the ion-sphere radius.
However, no cutoff procedure can fix the fact that Eq.~\eqref{eq:dcs-approx} is an unrealistic cross-section for dense plasmas.
If the qLFP theory is to be used as a quantitative theory of transport properties of dense plasmas, one should instead base the Coulomb logarithms on cross-sections that contain the relevant high-density physics -- diffraction, screening, exchange, and correlations -- rather than trying to insert these effects \textit{ad hoc}.
This is accomplished by treating the scattering physics quantum mechanically, which naturally incorporates diffraction and exchange, as well as by choosing an appropriate effective scattering potential.
The question is then what potential should be used to ensure these effects are adequately modeled.

Over the past several years, \emph{mean-force} scattering has proved to be a useful concept for describing transport in correlated plasmas within a binary collision kinetic theory\cite{BaalrudPRL2013,DaligaultPRL2016,ShafferPRE2017,StarrettHEDP2017,ShafferPOP2019}.
The principle is that scattering between particles in a plasma should be described not by the Coulomb interaction but by the potential of mean force, $V_{ij}^\mf(r)$.
The potential of mean force corresponds to the effective force between two particles one obtains by fixing their positions a distance $r$ apart and canonically averaging over all configurations of the remaining particles of the plasma.
In a weakly coupled plasma, $V^\mf_{ij}(r)$ recovers the Debye-H\"uckel potential
\begin{equation}
  V^\mf_{ij}(r) \to \frac{e_i e_j}{r} e^{-\kappa r}
\end{equation}
where $\kappa$ is the Debye wave number\cite{HansenMacDonald}.
In strongly coupled plasmas, $V^\mf_{ij}(r)$ reflects the onset of short-ranged order in the plasma.
For electron transport, one needs only the electron-ion and electron-electron mean-force potentials, which we obtain from an average atom model as described in Refs.~\cite{StarrettHEDP2017} and \cite{ShafferPRE2020} respectively, as well as references therein.
\comment{This average-atom model also provides the effective ion charge, $Z$, and the electron chemical potential, $\mu_e$.
They are related by
\begin{equation}
  n_e = 2 {\left( \frac{m_ek_BT}{2\pi\hbar^2} \right)}^\frac32 \qint_{\frac12}(\beta\mu_e)
  = Z n_I = Z \rho / m_I
\end{equation}
where $Q_{\frac12}$ is a Fermi-Dirac integral defined by
\begin{align}
  \label{eq:qint}
  & \qint_\nu(z) = \frac{1}{\Gamma(\nu+1)}\int_0^\infty \frac{x^\nu}{e^{x-z} + 1} dx 
\end{align}
Details on the average atom model may be found in Ref.~\cite{StarrettPRE2013}, where the relevant ionization, electron density, and chemical potential are notated as $\bar Z$, $\bar n_e^0$, and $\mu_e^{\mathrm{id}}$ respectively.}

The electron-ion and electron-electron mean force potentials are used to obtain the transport cross-section relevant to electron-ion collisions and electron-electron collisions, respectively.
First, radial Schr\"odinger equations are solved for scattering state wave functions
\begin{equation}
  \left[\dd{^2}{r^2} - \frac{l(l+1)}{r^2} - \frac{2m_{Ie}}{\hbar^2}V^\mf_{eI}(r) + k^2\right] P^I_{kl}(r) = 0
\end{equation}
\begin{equation}
  \left[\dd{^2}{r^2} - \frac{l(l+1)}{r^2} - \frac{m_e}{\hbar^2}V^\mf_{ee}(r) + k^2\right] P^e_{kl}(r) = 0
\end{equation}
to obtain the scattering phase shifts, $\delta^I_l(k)$ and $\delta^e_l(k)$.
For electron-ion scattering, we evaluate the momentum-transfer cross-section,
\begin{equation}
  \sigma^{(1)}_{Ie}(k) = \frac{4\pi}{k^2} \sum_{l=0}^\infty (l+1) \sin^2(\delta^I_{l+1} - \delta^I_l)
\end{equation}
and for electron-electron scattering, we evaluate the viscosity cross-section,
\begin{equation}
  \sigma^{(2)}_{ee}(k) = \frac{4\pi}{k^2} \sum_{l=0}^\infty \frac{(l+1)(l+2)}{2l+3} \sin^2(\delta^e_{l+2} - \delta^e_l) \left[ 1 - \frac12(-1)^l \right]
\end{equation}
taking care not to neglect that electrons are indistinguishable\cite{SpringerAMOHandbook}.

It is perhaps not obvious why the viscosity cross-section should be used for electron-electron scattering.
One reason is that the momentum-transfer cross-section gives \comment{high-$k$ behavior that is inconsistent with the qLFP equation.}
Consider a Debye-H\"uckel potential for electron-electron scattering.
In the first Born approximation, one finds for the differential, momentum-transfer, and viscosity cross-sections\footnote{
  It is interesting to note that for indistinguishable particles in general, the odd cross-sections, $\sigma_{ee}^{(1)}$, $\sigma_{ee}^{(3)}$, etc., reduce to the total cross-section, $\sigma_{ee}^{(0)}$. This is because the differential cross-section for indistinguishable particles an even function of $\cos\theta$, so odd powers of $\cos\theta$ in Eq.~\eqref{eq:sig-r} vanish.
}.
\begin{equation}
  \dd{\sigma_{ee}}\Omega \approx \frac{\pi a_B^2}{2}
  \frac{
    12 k^4\cos^2\theta + (\kappa^2 + 2k^2)^2
  }{
    [ 4k^4\cos^2\theta - (\kappa^2 + 2k^2)^2 ]^2
  }
  \label{eq:ee-dcs-dh}
\end{equation}
\begin{align}
  \sigma^{(1)}_{ee}(k)
  & \approx \frac{\pi a_B^2}{k^2} \left[ \frac{4k^2}{\kappa^2(\kappa^2 + 4k^2)} - \frac{\ln\left(1 + \frac{4k^2}{\kappa^2}\right)}{2\kappa^2 + 4k^2}\right]
    \label{eq:ee-1-dh}
  \\
  & \sim \frac{\pi a_B^2}{k^4} \left[ \frac{k^2}{\kappa^2} - \frac12\ln(2k/\kappa) - \frac14 + \bigO(k^{-1}) \right]
    \label{eq:ee-1-dh-hik}
\end{align}
\begin{align}
  \sigma^{(2)}_{ee}(k)
  & \approx \frac{\pi a_B^2}{k^4} \left[ \frac{16k^4 + 20\kappa^2k^2 + 5\kappa^4}{16k^4 + 8\kappa^2k^2}\ln\left( 1 + \frac{4k^2}{\kappa^2}\right) - \frac52 \right]
    \label{eq:ee-2-dh}
  \\
  & \sim \frac{\pi a_B^2}{k^4} \left[ 2\ln(2k/\kappa) - \frac52 + \bigO(k^{-1}) \right]
    \label{eq:ee-2-dh-hik}
\end{align}
Recalling that the qLFP theory assumes cross-sections with Rutherford-like behavior, it is clear that the asymptotic $k^{-2}$ dependence of the momentum-transfer cross-section is unsuitable, whereas the $k^{-4}\ln k$ dependence of the viscosity cross-section is exactly the prescribed scaling.
The choice to use the viscosity cross-section is further supported in the high-$T$ limit.
There, the Chapman-Enskog solution of the classical Boltzmann equation shows that $\sigma_{ee}^{(2)}$ is the electron-electron transport cross-section which appears in the transport coefficients\cite{FerzigerKaper}.

From the cross-sections, $\sigma_{Ie}^{(1)}$ and $\sigma_{ee}^{(2)}$, it is necessary to construct Coulomb logarithms.
To this end, it is useful to write them in the form
\begin{equation}
  \label{eq:sig-ie-pieces}
  \sigma_{Ie}^{(1)}(k) = \bar\sigma_{Ie}(k) \ln \Lambda_{Ie}(k)
\end{equation}
\begin{equation}
  \label{eq:sig-ee-pieces}
  \sigma_{ee}^{(2)}(k) = \bar\sigma_{ee}(k) \ln \Lambda_{ee}(k)
\end{equation}
where $\bar\sigma_{ij}(k) = \pi (e_ie_j m_{ij}/\hbar^2 k^2)^2$ is a reference Coulomb-like cross-section.
In the qLFP theory, the Coulomb logs are to be taken as constants, but clearly they will not be in general.
A procedure for reducing the residual $k$ dependence of the cross-sections into constant Coulomb logarithms is needed.

Electron-ion scattering is expected to be the dominant collision process at temperatures below the Fermi energy.
At low temperatures, it is also expected that the small-angle approximation will break down.
In order to get the best possible accuracy at low temperatures, we construct an electron-ion Coulomb logarithm based on a mean-force relaxation time approximation\cite{StarrettHEDP2017,StarrettPOP2018}.
This is a separate kinetic theory which ignores the effects of electron-electron collisions, but does not require a small-angle approximation (see Appendix~\ref{sec:trx-uu} for a derivation).
It recovers the transport coefficients for a Lorentz gas at high temperatures and is in good agreement with available QMD data on electrical conductivity at low temperatures where electron-electron scattering is expected to be unimportant\cite{StarrettHEDP2017}.
We construct a Coulomb logarithm by considering the electrical conductivity in the relaxation time approximation \comment{(corresponding to the ``direct'' averaging approach in the language of Ref.~\cite{BurrillHEDP2016})}
\begin{equation}
  \label{eq:sig-rt}
  \sigma_{RT} = -\frac13 e^2\int \tau(\epsilon) v^2 \pp{f_e^{(0)}}{\epsilon} \frac{d\vec p}{\omega_e}
\end{equation}
where $\tau^{-1} = n_I v \sigma_{Ie}^{(1)}$ is an energy-dependent relaxation time.
If the momentum-transfer cross-section is instead approximated by Eq.~\eqref{eq:sig-ie-pieces} with a constant Coulomb logarithm, one finds that the electrical conductivity simplifies to
\begin{equation}
  \label{eq:sig-cl}
  \sigma_{RT} \approx  \frac{32}{3\pi} \frac{\qint_2(\beta\mu_e)}{\qint_\frac12(\beta\mu_e)} \frac{n_e\bar\tau}{m_e} 
\end{equation}
with the mean relaxation time
\begin{equation}
  \label{eq:taubar}
  \bar\tau = \frac{3}{4\sqrt{2\pi}} \frac{m_e^\frac12 (k_BT)^\frac32}{Z e^4 n_e \ln\Lambda_{Ie}}
\end{equation}
\comment{We define the electron-ion Coulomb logarithm by requiring that Eq.~\eqref{eq:sig-cl} and Eq.~\eqref{eq:taubar} together reproduce Eq.~\eqref{eq:sig-rt} as a function of density and temperature,
\begin{equation}
  \label{eq:lnL-ei}
 \ln \Lambda_{Ie} = \frac{2^{\frac52}}{\pi^{\frac32}} \frac{Q_2(\beta\mu_e)}{Q_{\frac12}(\beta\mu_e)}
  \frac{(k_BT)^{\frac32}}{ Z m_e^{\frac12} \sigma_{RT}}
\end{equation}
which essentially just recasts the relaxation-time approximation into an effective Coulomb logarithm, similar to the construction of Ref.~\cite{StarrettPOP2018}.}
In this way, we are assured that the electron-ion scattering contributions to the transport coefficients will agree with the mean-force relaxation time approximation, which is known to be accurate at temperatures well below the expected range of validity of qLFP\cite{StarrettHEDP2017}.

For electron-electron scattering, we note the appearance of a logarithm in Eq.~\eqref{eq:ee-2-dh-hik}.
We construct an electron-electron Coulomb logarithm by rearranging this expression for the logarithm and averaging over $k$, 
\begin{equation}
  \label{eq:lnL-ee}
  \ln\Lambda_{ee} = \frac12 \ave{\frac{\sigma_{ee}^{(2)}(k)}{\bar\sigma_{ee}(k)}} + \frac54
\end{equation}
The average is with respect to the distribution of relative momenta of two electrons
\begin{equation}
  \ave{A} = \int_0^\infty A(x) F_{ee}(x) dx
\end{equation}
where $x = \hbar k/\sqrt{(m_e/2)k_BT}$ is the dimensionless relative momentum and
\begin{multline}
  F_{ee}(x)
  = \frac4\pi |\qint_\frac12(\beta\mu_e)|^{-2} \\
  \times \int_0^\infty \ln\left|\frac{1 + e^{\beta\mu_e-(x-y)^2}}{1 + e^{\beta\mu_e-(x+y)^2}}\right|\frac{xy \,dy}{e^{y^2 - \beta\mu_e} + 1} 
\end{multline}
Because this construction is based on the analytic form of the Born approximation with a simple potential, it is suitable for high temperatures, where electron-electron scattering is expected to be most important.
However, we note that at low temperatures, the second term in Eq.~\eqref{eq:lnL-ee} will artificially dominate the value of $\ln\Lambda_{ee}$, so that Eq.~\eqref{eq:lnL-ee} should be modified.
In Sec.~\ref{sec:results} it will be shown that rolling off the offset according to
\begin{equation}
  \label{eq:lnL-ee-mod}
  \ln\Lambda_{ee} = \frac12 \ave{\frac{\sigma_{ee}^{(2)}(k)}{\bar\sigma_{ee}(k)}} + \frac54 \comment{\erf{\left[{(2T/3T_F)}^3\right]}}
\end{equation}
produces substantially improved low-temperature behavior.
\comment{The error-function rolloff adopted here is just one of many physically plausible functional forms.
We limit our scope to this one only for definiteness, not because it holds any special physical significance.}

\begin{figure}
  \centering
  \includegraphics[width=\columnwidth]{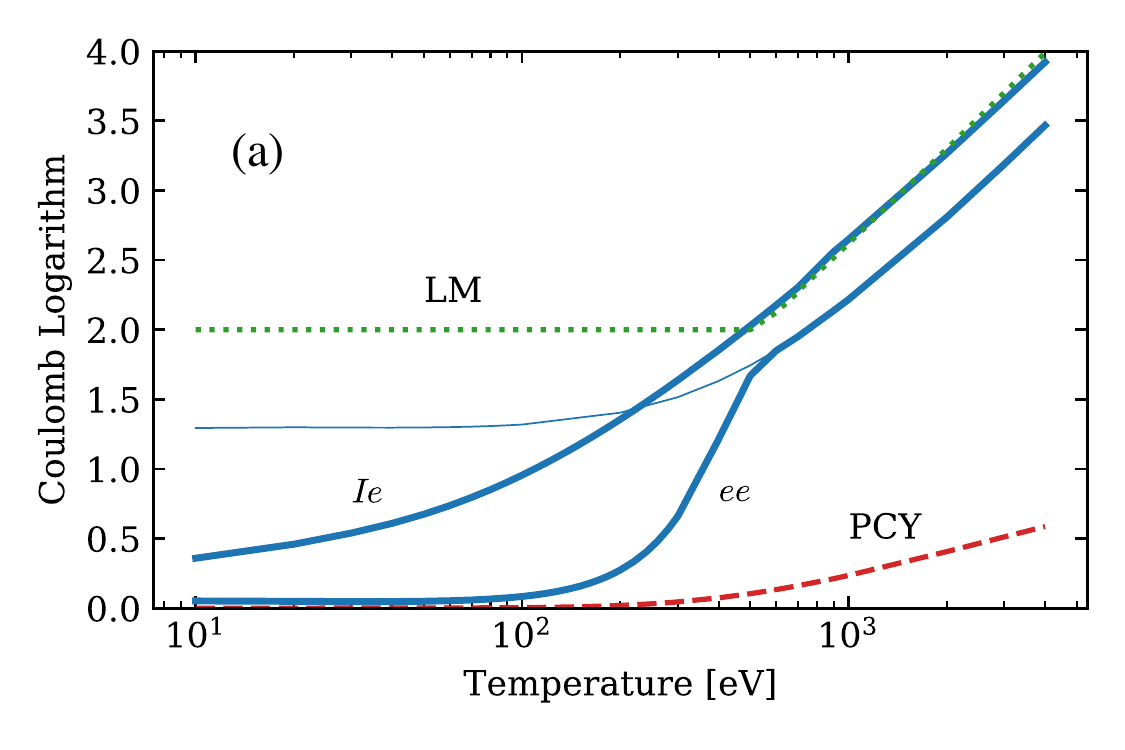}\vspace{-0.1in}
  \includegraphics[width=\columnwidth]{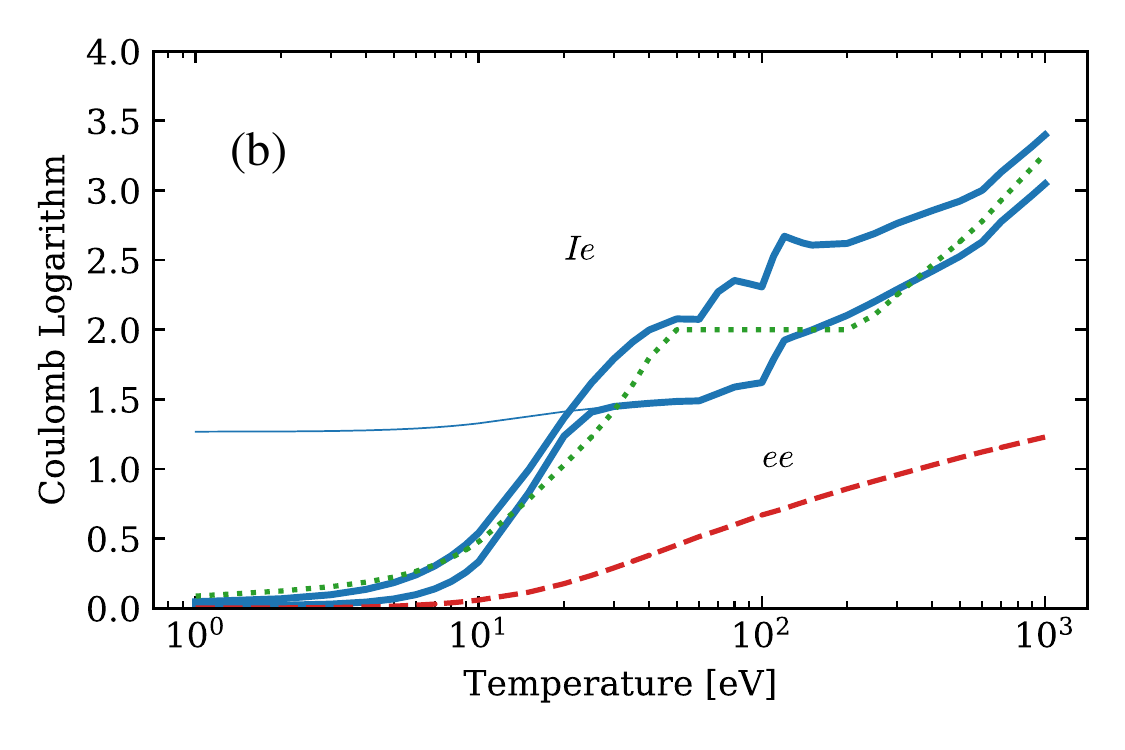}
  \caption{Coulomb logarithms for (a) compressed hydrogen at \SI{40}{\gram\per\cubic\centi\meter} and (b) solid-density aluminum at \SI{2.7}{\gram\per\cubic\centi\meter}. Thick solid curves are the mean-force model. The thin solid curve is the electron-electron mean-force model without low-temperature correction, Eq.~\eqref{eq:lnL-ee} versus Eq.~\eqref{eq:lnL-ee-mod}. The green dotted curve (LM) is the Lee-More Coulomb logarithm, Eq.~\eqref{eq:lnL-LM}, with their additional mean free path correction, Eq.~\eqref{eq:lnL-LM-mfp}. The red dashed curve is the Coulomb logarithm of Potekhin et. al, Eq.~\eqref{eq:lnL-pcy}.}
  \label{fig:lnL}
\end{figure}

Figure~\ref{fig:lnL} shows Coulomb logarithms for the two materials considered in Sec.~\ref{sec:results}, highly compressed hydrogen and solid-density aluminum.
The mean-force model recovers the expected logarithmic temperature scaling at high temperature without needing the \textit{ad hoc} Debye screening cutoffs used in traditional theory.
At lower temperatures, non-logarithmic behavior is captured without needing to patch together different physical models as in the Lee-More Coulomb logarithm.
The ``wiggles'' in the aluminum $\ln\Lambda_{Ie}$ are also due to the mean-force potential and are discussed further in Sec.~\ref{sec:solidAl}.
For electron-electron scattering, we also see the effect of the low-temperature correction proposed in Eq.~\eqref{eq:lnL-ee-mod}; rather than $\ln\Lambda_{ee}\to\frac54$ (thin curve in Fig.~\ref{fig:lnL}), the modified Coulomb logarithm rapidly decreases, which is more physically reasonable behavior.
\comment{For comparison, we also show the electron-electron Coulomb logarithm implied by the practical formula by Potekhin et.~al~\cite{PotekhinAA1997} for the electron-electron collision rate in degenerate plasmas.
  In their notation, we obtain
  \begin{equation}
    \label{eq:lnL-pcy}
    \ln\Lambda^{\mathrm{PCY}}_{ee} = \frac{5x^4}{2\sqrt{3}y^3} J(x, y)
  \end{equation}
  where $x=p_F/m_ec$, $y=\sqrt{3}\hbar\omega_{pe}/k_BT$, $p_F = \hbar(3\pi^2n_e)^{\frac{1}{3}}$ is the Fermi momentum, $\omega_{pe}=\sqrt{4\pi n_ee^2/m_e}$ is the plasma frequency, $c$ is the speed of light, and $J(x, y)$ is a dimensionless quantity given by Eq.~(A3) of Ref.~\cite{PotekhinAA1997}.
  The Coulomb logarithm is identified by matching the nonrelativistic and static-screening limit ($x,y\ll1$) of Potekhin et.~al's formula to the analytic expression for the electron-electron contribution to the thermal conductivity derived by Lampe, Eq.~(16) of Ref.~\cite{LampePR1968a}.
  Since the model of Ref.~\cite{PotekhinAA1997} is not intended for high temperatures, its behavior in this regime is not correct, with the e-e Coulomb logarithm being far too small in the classical limit.
  At low temperatures, where their model is indended to be used, the Coulomb logarithm rapidly approaches zero, i.e., that electron-electron scattering ceases to influence transport.
This limit is approached somewhat more rapidly than in the present model.}

For ion-ion scattering, no special model for the Coulomb logarithm is necessary.
The values of the electron transport coefficients are insensitive to the ion-ion collision physics.
To a good approximation, one can take the ions to be in local thermodynamic equilibrium on the time scales relevant to electron hydrodynamics, so that the ion-ion collision operator is identically zero.
Operationally, this can be achieved by setting $\ln\Lambda_{II}=0$.
In our calculations, we find no meaningful difference between zeroing the ion-ion Coulomb logarithm versus using the model of Brysk et al.\cite{BryskPP1975}.

\section{Results}
\label{sec:results}

In this section we present results for the electron transport coefficients from the Chapman-Enskog solution of the linearized qLFP kinetic equation, detailed in Ref.~\cite{DaligaultPOP2018} (with some minor corrections listed in Appendix~\ref{app:typo}).
See Appendix~\ref{app:ce} for a condensed practical discussion of the method\comment{, as well as formulas for the electrical conductivity [Eq.~\eqref{eq:sigma}] and thermal conductivity [Eq.~\eqref{eq:lambda}].}
Our solution expands the distribution functions in Daligault's polynomial basis, which guarantees optimal and monotonic convergence with respect to the basis size.
The lowest degree polynomials coincide (up to constant factors) with those used by Lampe\cite{LampePR1968b}.
We find that for degenerate conditions, two polynomials are generally sufficient to converge the electrical and thermal conductivities within one percent, but that nondegenerate conditions typically require three polynomials.
The rapid convergence of Daligault's polynomial basis at arbitrary degeneracy is the main technical benefit of the Chapman-Enskog approach.
We would expect similar benefits if Daligault's polynomials were used as relevant observables in the Zubarev linear response approach\cite{RopkePRA1988,ReinholzPRE2015,RopkeBook}, which could alleviate the reported convergence difficulties at intermediate temperatures\cite{OvechkinHEDP2016}.
The electrical and thermal conductivities shown here are computed in the five- and four-polynomial approximations respectively.
We have also computed the thermoelectric coefficient and have found it to be less sensitive to the detailed collision physics than the conductivities.
For this reason and to keep the discussion focused, we do not show our results for the thermoelectric coefficient.

We consider two prototypical cases below: highly compressed hydrogen as well as solid-density aluminum.
Hydrogen is singled out for its importance in ICF and astrophysical modeling, as well as it being the material which is most sensitive to electron-electron collisions.
Aluminum provides an interesting contrast because it is partially ionized at the conditions considered and has relatively simple, but nontrivial, shell structure which is strongly temperature dependent.
\comment{We evaluate the qLFP electrical and thermal conductivities using Eqs.~\eqref{eq:sigma} and~\eqref{eq:lambda} respectively, taking $\ln\Lambda_{Ie}$ from Eq.~\eqref{eq:lnL-ei} and $\ln\Lambda_{ee}$ from Eq.~\eqref{eq:lnL-ee-mod}.
We also show the qLFP results when Eq.~\eqref{eq:lnL-ee} is used for $\ln\Lambda_{ee}$, that is, without low-temperature correction.}
We compare with available QMD data, as well as the analytic model by Lee and More\cite{LeePF1984} and the tables by Rinker\cite{RinkerPRB1985a,RinkerPRB1985b}.
When evaluating the Lee-More model, we use only their plasma model with the collision rate given in terms of the Coulomb logarithm, Eq.~\eqref{eq:lnL-LM}, except at very low temperatures where their model's mean free path $\bar\tau\sqrt{3k_BT/m_e} \qint_\frac12/\qint_{-1}$ is smaller than $a_I$, the Coulomb logarithm is replaced by
\begin{equation}
  \label{eq:lnL-LM-mfp}
  \ln\Lambda_{\mathrm{LM}} = \sqrt{\frac{3\pi}{2}} \left(\frac{a_Ik_BT}{Z^2e^2}\right)^{\!2} \frac{\qint_\frac12(\beta\mu_e)}{\qint_{-1}(\beta\mu_e)}
\end{equation}
which corresponds to replacing the mean free path with $a_I$ in their formulas.
We evaluate the model using the same average-atom ionization, electron density, and chemical potential as for the qLFP calculations.


\subsection{Compressed Hydrogen}
\label{sec:denseH}

\begin{figure*}[t]
  \centering
  \includegraphics[width=\linewidth]{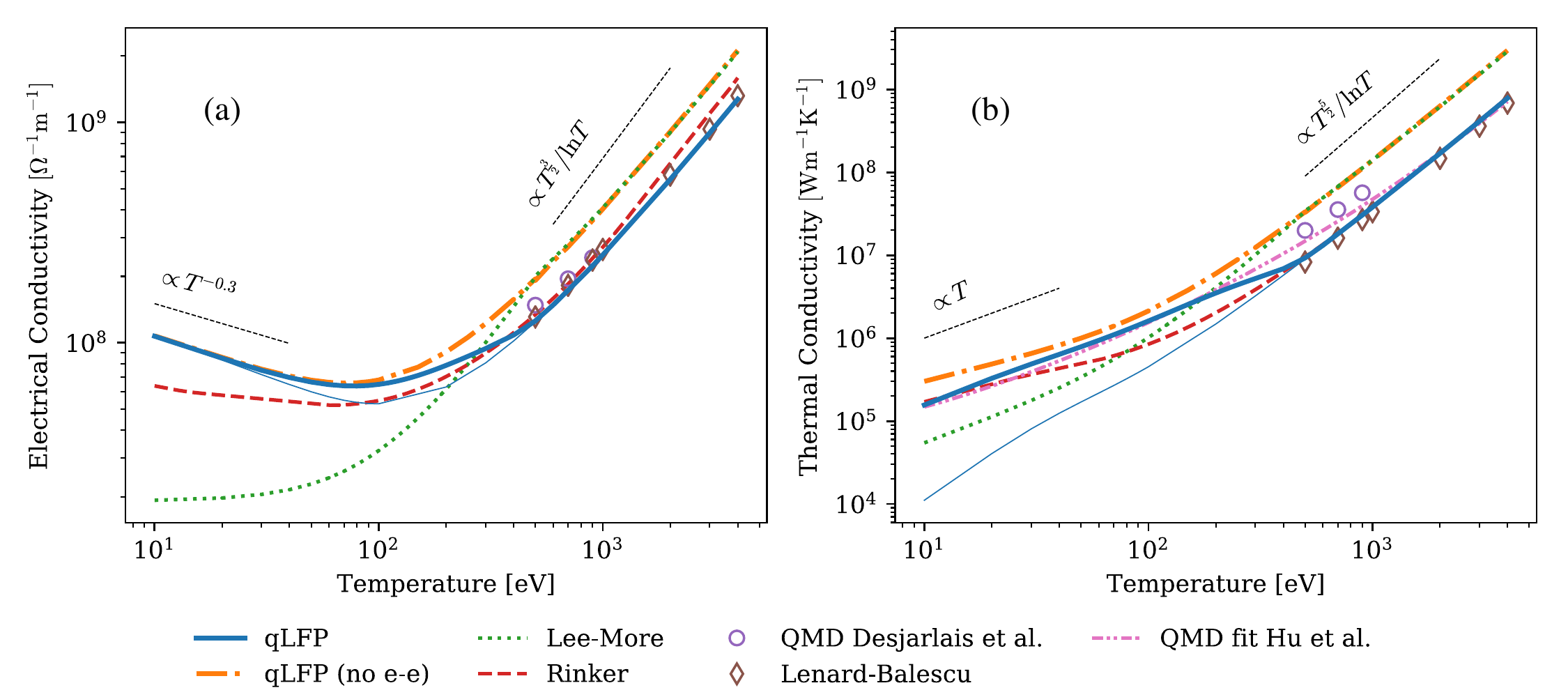}
  \caption{Electrical conductivity (a) and thermal conductivity (b) of hot dense hydrogen plasma. The thin and thick qLFP curves are with and without low-temperature correction, as in Fig.~\ref{fig:lnL}.}
  \label{fig:denseH} 
\end{figure*}

Our results for compressed hydrogen are shown in Figure~\ref{fig:denseH}.
The conditions considered are the \SI{40}{\gram\per\cubic\centi\meter} isochore from temperatures of \SI{10}{\electronvolt} to \SI{4}{\kilo\electronvolt}.
Throughout, the solid blue curves are our qLFP results, the dash-dotted orange curves are qLFP without electron-electron scattering, the dotted green curves are the Lee-More plasma model, and the red dashed curves are an interpolation of Rinker's tables.
At high temperatures, we compare with the QMD and Lenard-Balescu results of Desjarlais et al.\cite{DesjarlaisPRE2017} for the electrical and thermal conductivity.
The deuterium thermal conductivity model by Hu et al.\cite{HuPRE2014b} is also shown, with the mass density appropriately scaled for hydrogen.
The thin-set qLFP curves represent the Coulomb logarithms and transport coefficients obtained when $\ln\Lambda_{ee}$ is given by Eq.~\eqref{eq:lnL-ee}, i.e., not corrected at low temperatures.

At \SI{40}{\gram\per\cubic\centi\meter}, the transition from classical to Fermi statistics happens in the vicinity of $T=\SI{300}{\electronvolt}$, where $T\approx T_F$.
Well above this temperature, the electrons are nondegenerate, the plasma is weakly coupled, and all models recover the qualitative scaling of the transport coefficients known from classical plasma theory\cite{SpitzerBook}
\begin{equation}
  \sigma \propto T^\frac32 / \ln T \qquad \lambda \propto T^\frac52 / \ln T 
\end{equation}
which can be derived from dimensional analysis assuming a Coulomb cross-section.
At high temperatures, we also find good quantitative agreement between the qLFP theory and Desjarlais et al.'s semiclassical Lenard-Balescu calculations, the differences being only a few percent.
This is because at high temperatures, the potential of mean force used in the qLFP calculations becomes the Debye-H\"uckel potential, in which case the qLFP collision operator is the same as the static screening limit of Lenard-Balescu.
Dynamic screening makes only a small, constant correction to the Coulomb logarithm at high temperature\cite{GouldPR1967,RopkePRA1988,HohnePA1984,BrownPR2005,*BrownPRE2009,DaligaultPRE2009}, which explains the agreement between qLFP and Lenard-Balescu in this regime.

\comment{The only source of major discrepancy between predictions at high temperature is the treatment of electron-electron scattering.
There are two ways in which electron-electron scattering can influence transport.
The first is the direct (or ``explicit'' in the words of Ref.~\cite{DesjarlaisPRE2017}) transport of energy (but not momentum) via electron-electron collisions from one part of the plasma to another.
The second is the indirect role of electron-electron scattering  in determining the shape of the steady-state electron distribution function in response to an applied electric field and/or temperature gradient.
The thermal conductivity is affected by both mechanisms, while the electrical conductivity is affected only by the indirect reshaping effect.
Models which neglect electron-electron scattering entirely, e.g., Lee-More or the ``no e-e'' qLFP results in Fig.~\ref{fig:denseH}, essentially predict the transport coefficients of a Lorentz gas.
A Lorentz gas model of hydrogen is known to predict an electrical conductivity that is about a factor of two too large and a thermal conductivity a factor of four too large at high temperature; Spitzer and H\"arm's classical results are 1.719 and 4.241, respectively~\cite{SpitzerPR1953}, which are close to the values of our numerical qLFP results (see Fig.~\ref{fig:denseH-reinholz} below).
The QMD calculations by Desjarlais et al.~predict an electrical conductivity that is in good agreement with the qLFP and Lenard-Balescu theories, but a thermal conductivity that is roughly a factor of two too large compared with the kinetic theories.
When electron-electron collisions are dropped from the qLFP calculations (``no e-e''), the resulting electrical and thermal conductivities are each about a factor of two larger than the QMD results.
These findings support the conclusions of Ref.~\cite{DesjarlaisPRE2017} that the Kubo-Greenwood QMD calculations contain the indirect electron-electron reshaping effect relevant to both the electrical and thermal conductivity, but they do not contain the direct scattering effect which further reduces the thermal conductivity.}

The transition region from Maxwell-Boltzmann to Fermi-Dirac statistics occurs for temperatures below about \SI{500}{\electronvolt}.
Here, the scaling of the transport coefficients deviates significantly from the classical scaling as the characteristic electron energy scale transitions from the temperature to the chemical potential.
The transition region culminates in a minimum in the electrical conductivity around \SI{100}{\electronvolt}.
The cause of this minimum is discussed in Sec.~\ref{sec:cond-min}.
For low temperatures below the electrical conductivity minimum ($e^{\beta\mu_e} \gg 1$), the electrical conductivity approaches the value
\begin{equation}
  \sigma \to  \frac{2\sqrt{2}}{3\pi^2} \frac{e^2\sqrt{m_e}}{\hbar^3}  \mu_e^\frac32  \tau(\mu_e) 
\end{equation}
If the temperature dependence of $\tau$ is neglected, then one finds from a Sommerfeld expansion of the relaxation-time approximation that conductivity should decrease quadratically with temperature, $\sigma \propto T^{-2}$\footnote{
  The Sommerfeld expansion gives  $\sigma(T)\sim \sigma(0) - c T^2$, which can be renormalized to $\sigma(0)(1 + cT^2)^{-1}$ to the same order of approximation.
}.
However, we find that the hydrogen conductivity approaches the minimum significantly slower than this (empirically, about $T^{-0.3}$) owing to fact that the cross-section actually has non-trivial temperature dependence through the potential of mean force.

The thermal conductivity at low temperatures is observed to scale roughly proportional to $T$, which is the scaling predicted by theories which neglect electron-electron scattering.
While all thermal conductivity models shown in Fig.~\ref{fig:denseH} roughly follow this scaling, the treatment of electron-electron scattering can make order-of-magnitude differences in the value of the thermal conductivity at low temperatures This sensitivity will be discussed further in Sec.~\ref{sec:ee-scat}.

The thermal conductivity fit by Hu et al.~warrants special mention.
The fit is constrained by QMD data at low temperatures ($T \lesssim T_F$) and Spitzer-type model at high temperatures ($T \gtrsim 3T_F$), but is unconstrained in between.
In Fig.~\ref{fig:denseH}, this range corresponds to $\SI{300}{\electronvolt} \lesssim T \lesssim \SI{900}{\electronvolt}$, where it is seen that Hu et al.'s interpolation overestimates the thermal conductivity relative to qLFP and Lenard-Balescu (for visual reference, recall that the Desjarlais et al.~QMD data is about a factor of two larger than the theoretical models).
Now that reliable theoretical models in this regime are available, interpolative models such as Hu et al.'s can be systematically improved for hot dense plasma conditions.


\subsection{Solid Density Aluminum}
\label{sec:solidAl}

\begin{figure*}[t]
  \centering
  \includegraphics[width=\linewidth]{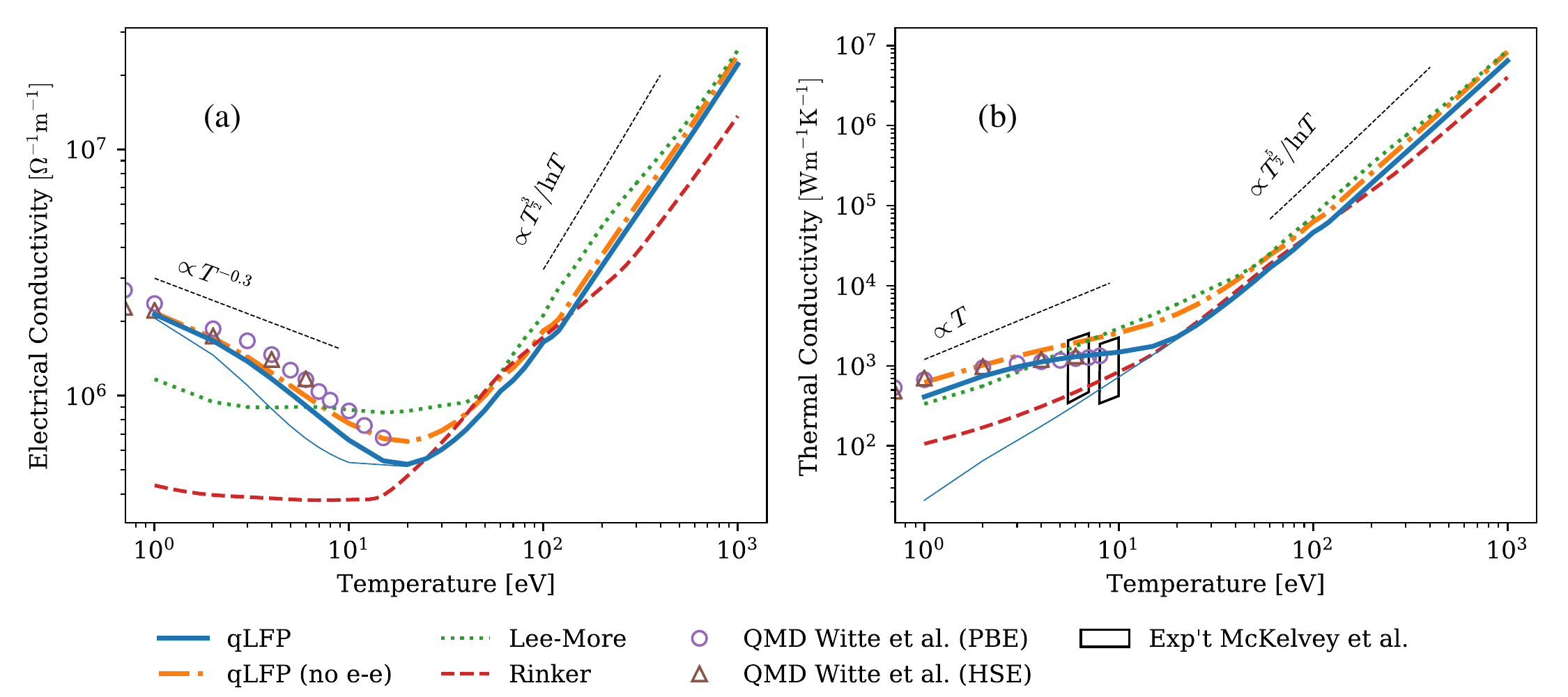}
  \caption{Electrical conductivity (a) and thermal conductivity (b) of solid-density aluminum plasma ($\rho=\SI{2.7}{\gram\per\cubic\centi\meter}$). The thin and thick blue curves are qLFP curves with and without low-temperature correction, as in Fig.~\ref{fig:lnL}. \comment{The dotted green curves are the Lee-More model~\cite{LeePF1984}. The dashed red curves are interpolations of Rinker's tables~\cite{RinkerPRB1985a,RinkerPRB1985b}. The circles and triangles are QMD data by Witte et al.~\cite{WittePOP2018} using PBE and HSE exchange-correlation functionals respectively. The black outlined regions are estimated ranges of thermal conductivity measured by McKelvey et al.~\cite{McKelveySR2017}.}}
  \label{fig:solidAl} 
\end{figure*}

Our results for solid-density ($\rho=\SI{2.7}{\gram\per\cubic\centi\meter}$) aluminum are shown in Figure~\ref{fig:solidAl}.
Some qualitative behaviors of the transport coefficients are similar to those seen in hydrogen.
For instance, aluminum also exhibits a minimum electrical conductivity, and the temperature scalings on either side is similar to those found for hydrogen, with the exception that the low-temperature behavior of the electrical conductivity seems somewhat more complex.
There are, however, two important ways in which aluminum differs from hydrogen.

\begin{figure}[b]
  \centering
  \includegraphics[width=\columnwidth]{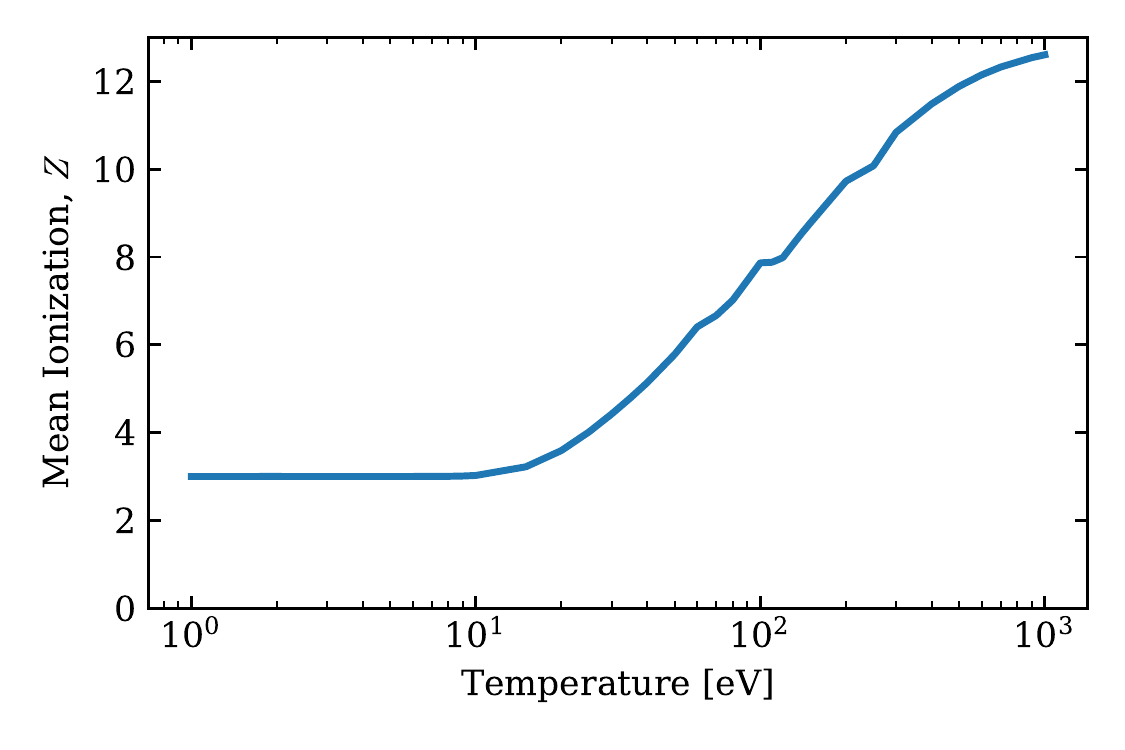}
  \caption{Mean ionization state of aluminum at a density $\rho = \SI{2.7}{\gram\per\cubic\centi\meter}$ as predicted by the average-atom model of Ref.~\cite{StarrettPRE2013}.}
  \label{fig:solidAl-zbar}
\end{figure}

The first is that over the density and temperatures investigated, the ionization state of aluminum strongly varies, there being three conduction electrons per atom ($Z=3$) below about \SI{10}{\electronvolt}, which steadily increases with temperature to nearly full ionization ($Z=13$) at \SI{1}{\kilo\electronvolt}.
\comment{The aluminum mean ionization predicted by the average-atom model of Ref.~\cite{StarrettPRE2013} is plotted in Fig.~\ref{fig:solidAl-zbar}, showing a mostly smooth transition from $Z=3$ at low temperature to near complete ionization at high temperature, with occasional starts and stops in between due the ionization of subshells.}
It is known that for classical plasmas, the influence of electron-electron scattering on the electrical conductivity is $\bigO(Z^{-1})$ compared to that of electron-ion scattering\cite{BraginskiiJETP1958,SimakovPOP2014}.
This is a consequence of the scaling of the collision operator as
\begin{equation}
  \label{eq:collop-scaling}
  C_{ij} \propto e_i^2e_j^2 n_i n_j \ln\Lambda_{ij}
\end{equation}
which holds true for the qLFP operator as well at high temperature.
The large ion charge at high temperature leads to electron-electron scattering being relatively unimportant for hot aluminum.
At low temperature, one expects that Pauli blocking should further suppress electron-electron collisions relative to electron-ion ones; however, this is not the case for qLFP for reasons discussed later in Sec.~\ref{sec:ee-scat}.

The second way in which our aluminum predictions differ qualitatively from hydrogen is that the binary scattering physics is much more complicated, owing both to the shell structure of partially ionized aluminum and stronger electron-ion coupling compared to hydrogen due to the higher ionic charge.
The combined effects of effects partial ionization and strong coupling result in significant non-monotonicity in the aluminum electron-ion mean-force potential, leading to resonant scattering.
If the electron-ion potential is monotonic (as is the case for hydrogen at most temperatures), then each angular momentum channel can support at most a single resonance from the interplay between the centrifugal barrier and the pair interaction.
However, when the mean-force potential has local maxima due to ionic correlations and/or shell structure, each of these can create additional centrifugal barriers and thus additional resonances.
The influence of resonant scattering is strongly temperature-dependent because each resonance only occurs over a narrow band of energies.
In order for a particular resonance to contribute to transport, its energy band must coincide with energies that are substantially occupied and not blocked by the exclusion principle, i.e., where $f_e^{(0)}[1 - f_e^{(0)}]$ is non-vanishing, viz. Eq.~\eqref{eq:sig-rt}.
The overall effect is clearest in the electron-ion Coulomb logarithm, plotted in Fig.~\ref{fig:lnL}b, which exhibits local minima and maxima between \SIrange{50}{150}{\electronvolt}, over which successive resonances are emphasized and then de-emphasized by the thermal distribution of electrons.
These resonant scattering features in $\ln\Lambda_{eI}$ are not as dramatic in the transport coefficients.
This is in part due to the logarithmic scale in the plots, but also because small variations in the Coulomb logarithm are swamped out by the stronger algebraic temperature scaling of the electrical and thermal conductivity ($T^\frac32$ and $T^\frac52$ respectively).

We also note that the electron-electron Coulomb logarithm shows a significant drop between \SIrange{30}{100}{\electronvolt}, which is similarly due to non-monotonicities in the electron-electron mean-force potential caused by the indirect influence of strong ion coupling.
However, this feature, like electron-electron scattering generally for aluminum, does not influence the transport coefficients in this temperature range.

\begin{figure}
  \centering
  \includegraphics[width=\columnwidth]{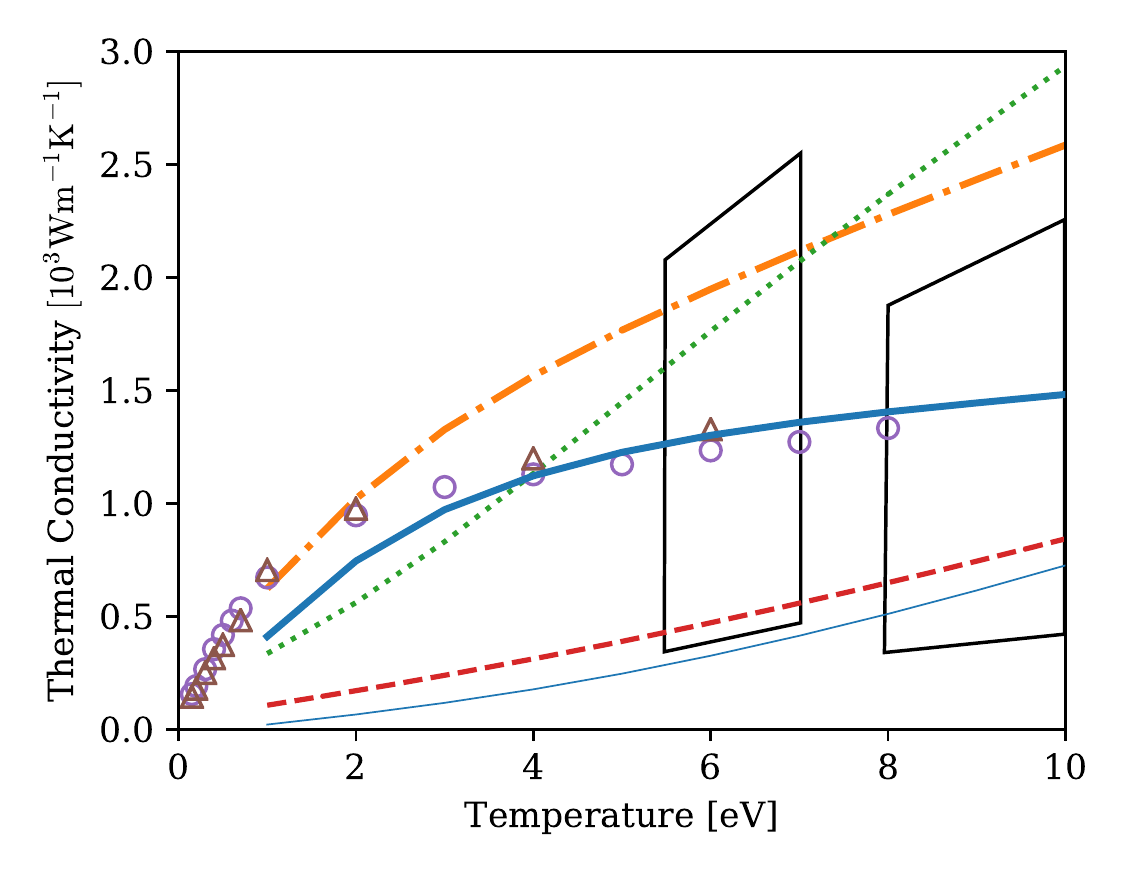}
  \caption{Low-temperature behavior of the thermal conductivity of solid-density aluminum. Curves and symbols have the same meaning as in Fig.~\ref{fig:solidAl}.}
  \label{fig:solidAl-zoom}
\end{figure}

The qLFP results for the electrical conductivity and thermal conductivity are compared with the values obtained in two sets of QMD simulations by Witte et al.~\cite{WittePOP2018}, the Lee-More model, Rinker's model, and experimental estimates by McKelvey et al~\cite{McKelveySR2017}.
We find that at the low temperatures where QMD data is available, the electrical conductivity is slightly more accurately predicted by qLFP when electron-electron collisions are neglected.
For the thermal conductivity, the inclusion of electron-electron collisions gives better agreement with the QMD data above \SI{3}{\electronvolt}, but below this qLFP compares better when electron-electron collisions are neglected, see Fig.~\ref{fig:solidAl-zoom}.
The reason for this has to do with the breakdown of the small-angle approximation at low temperature, which leads qLFP to over-emphasize the importance of electron-electron collisions, discussed further in Sec.~\ref{sec:ee-scat}.
Nevertheless, it is striking that even at temperatures as low as \SIrange{3}{10}{\electronvolt} (less than $T_F\approx \SI{12}{\electronvolt}$), the electron-electron scattering contribution to the thermal conductivity is important to include.
Evidently, Pauli blocking only extinguishes the influence of electron-electron scattering at very low temperatures.

Compared with Rinker's tables, qLFP produces significantly better predictions for the electrical and thermal conductivity below about \SI{30}{\electronvolt}.
Rinker's calculations suffer from a poor ionization model, which predicts for aluminum that $Z\to1$ as $T\to0$ instead of the physically correct $Z=3$.
This error leads to free electron densities and chemical potentials that are too small as well.
Since the low-temperature transport coefficients are sensitive to the value of the scattering cross-sections at the Fermi surface, an incorrect $Z$ in Rinker's tables translates to large errors in the transport coefficients, seen clearly in Fig.~\ref{fig:solidAl-zoom}.
The experimental results by McKelvey et al.~are able to rule out the Lee-More model in spite of the large uncertainty.
They also corroborate the QMD results by Witte et al.~and give some further confidence that qLFP produces the best predictions for the thermal conductivity of aluminum of the models considered, even at warm dense matter conditions.

\section{Discussion}
\label{sec:discussion}

\subsection{The Conductivity Minimum}
\label{sec:cond-min}

One cause of the predicted minimum in the electrical conductivity is the onset of spatial ordering in the ions; the plasma begins to take on characteristics of a liquid metal.
In liquid metals, the standard theory of conduction is based on the Ziman formula, which has been applied to warm and hot dense plasmas by many authors\cite{BurrillHEDP2016,RozsnyaiHEDP2008,PerrotPRA1987,SterneHEDP2007,PainCPP2010}.
In particular, Burrill et al.~demonstrated that the electrical conductivity minimum can only be captured by a Ziman-type theory when ionic correlations are accounted for.
Here, the qLFP results are based on a gas kinetic theory, but the ionic structure is accounted for in the mean-force scattering potential, so the electrical conductivity minimum is obtained.
In contrast, similar gas-kinetic models with analytic Coulomb logarithms, e.g., Lee-More, either do not capture the conductivity minimum (as for compressed hydrogen) or severely misplace it (as for solid aluminum) and are not suitable for this transitional temperature regime.

However, there is a simpler argument for why a conductivity minimum must occur based only on Pauli blocking.
At low temperatures, the electron mean free path elongates due to the exclusion principle.
Since an electron's energy changes very little when colliding with an ion, only electrons near the Fermi surface with energies $E \approx \mu_e \pm k_BT$ collide with ions.
Thus, with decreasing temperature, the fraction of electrons which can resist a current diminishes, so the conductivity must increase.
In this way of thinking, the conductivity minimum is just a necessary consequence of any model that captures correct qualitative behavior at both low and high $T$.

\subsection{Electron-Electron Scattering}
\label{sec:ee-scat}

In both the compressed hydrogen and solid-density aluminum cases considered in Sec.~\ref{sec:results}, qLFP predicts that at low temperatures, electron-electron collisions affect the thermal conductivity more than they do the electrical conductivity.
The inclusion or omission of electron-electron collisions in qLFP does not significantly affect the value of the electrical conductivity at low temperatures, whereas the thermal conductivity still depends on electron-electron collisions down to the lowest temperatures considered.
Analytically, this comes through in the explicit calculation of the qLFP electrical and thermal conductivities in the one-polynomial approximation and using $m_e\ll m_I$, (both of which are good approximations in this case)\cite{DaligaultPOP2018}\footnote{Compared with the expressions in Ref.~\cite{DaligaultPOP2018}, we have exploited some further analytic simplifications of the integrals $\mathcal A_{eI}^{p,q,r}$ for the neutral plasma electrical and thermal conductivity that lead to our Eqs.~\eqref{eq:degen-sigma-1} and~\eqref{eq:degen-lambda-1}.}
\begin{equation}
  \label{eq:degen-sigma-1}
  [\sigma]_1 = \frac{9\mu^\frac32}{16\pi\sqrt{2m_e} Ze^2\ln\Lambda_{Ie}} + \bigO(\beta\mu_e)^{-\frac52}
\end{equation}
\begin{equation}
  \label{eq:degen-lambda-1}
  [\lambda]_1 = \frac{5\pi^3k_B^3T^2\sqrt{\mu_e}}{36\sqrt{2m_e}e^4\ln\Lambda_{ee}} + \bigO(\beta\mu_e)^{-\frac32}
\end{equation}
It is seen that $[\sigma]_1$ depends only on the electron-ion collisions, while $[\lambda]_1$ depends only on the electron-electron collisions.
In fact, the thermal conductivity at low temperatures is identical with that of an electron gas\footnote{Note that in Ref.~\cite{DaligaultPOP2018}, the quoted expression for the electron gas thermal conductivity has inverted temperature dependence.}.
This result is at rather striking odds with the conventional theory of conduction in simple metals (solid and liquid), where one expects electron-electron collisions to be insignificant to both electrical and thermal conduction, which should be approximately related by the Wiedemann-Franz law\cite{ZimanFermiSurface}.

The reason for this has to do with the fact that Pauli blocking of electron-electron collisions in degenerate plasma is somewhat more nuanced than electron-ion collisions, a point first articulated by Lampe\cite{LampePR1968a}.
At low temperatures, an electron is only likely to participate in a collision if its pre- and post-collision energies lie within a range $E\approx \mu_e \pm k_BT$. 
For electron-ion collisions, the tiny change in the electron's energy after colliding means that only those electrons within this smeared-out Fermi surface suffer meaningful collisions.
For electron-electron collisions, both particles' energies are restricted to the vicinity of the Fermi surface.
Lampe's insight is that this condition implies large-angle electron-electron collisions are more strongly Pauli blocked than electron-ion ones, whereas small-angle collisions are less so\cite{LampePR1968a}.

We can then conclude that the persistent influence of electron-electron collisions in the thermal conductivity at low temperatures is an artifact of the small-angle approximation.
Plasmas at these conditions are not only degenerate but also strongly coupled and strongly screened, which leads to transport being controlled mainly by the low-energy and large-angle scattering part of the cross-sections.
Consequently, one should expect that electron-ion and electron-electron collisions should both be strongly Pauli blocked, but the electron-electron ones more so.
This would lead to both electrical and thermal conductivities being determined mainly by electron-ion collisions.
However, the small-angle approximation in qLFP changes things substantially for the reasons pointed out above and by Lampe.
The small-angle approximation does not change the degree to which electron-ion collisions are Pauli blocked, but it does weaken the Pauli blocking effect on electron-electron collisions so that both processes are about equally restricted.
For thermal conduction, qLFP then predicts that electron-electron scattering is the dominant process because it is a much more efficient means of changing individual electrons' energy than electron-ion collisions.
This approximate treatment of electron-electron Pauli blocking in qLFP means that the theory, while successful over a wide range of temperatures, does eventually break down for sufficiently degenerate plasmas.

\begin{figure}
  \centering
  \includegraphics[width=\columnwidth]{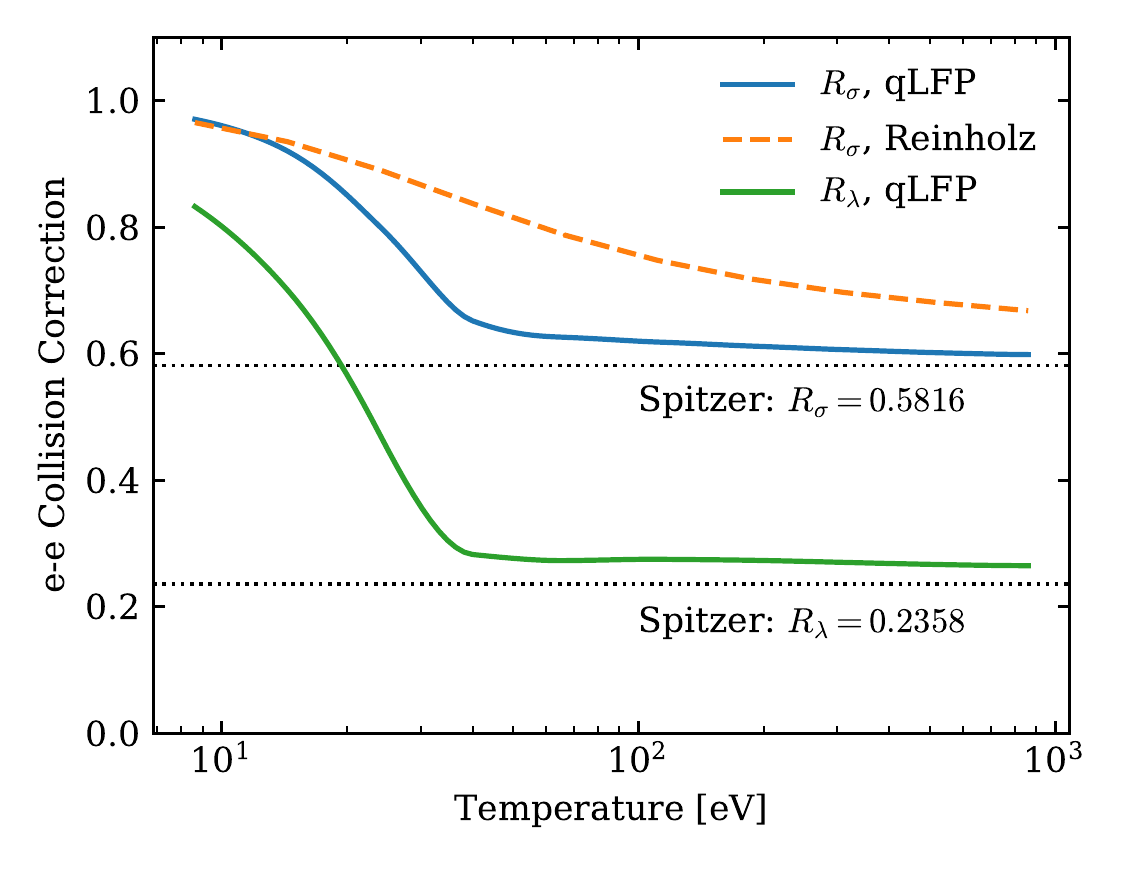}
  \caption{Electron-electron correction factor to the electrical conductivity, $R_\sigma$, and thermal conductivity, $R_{\lambda}$, for compressed hydrogen at $\rho = \SI{1}{\gram\per\cubic\centi\meter}$. Solid blue and green lines are the present qLFP model. The dashed orange line is the practical formula by Reinholz et al.~\cite{ReinholzPRE2015}. The dotted black lines are the Spitzer-H\"arm result~\cite{SpitzerPR1953}.}
  \label{fig:denseH-reinholz}
\end{figure}

\comment{%
The importance of electron-electron scattering in the electrical and thermal conductivity has also been quantitatively investigated in recent years by Reinholz et al.~\cite{ReinholzPRE2015} and Desjarlais et al.~\cite{DesjarlaisPRE2017}.
Ref.~\cite{ReinholzPRE2015} presents a practical formula for an electron-electron collision correction
\begin{equation}
  R_{\sigma} = \frac{\sigma}{\sigma(\text{no e-e})}
\end{equation}
which is the ratio of the electrical conductivity to that of a Lorentz plasma.
Their formula, which may be found in Eq.~(34) of Ref.~\cite{ReinholzPRE2015}, is based on the Zubarev linear response theory, with collision integrals evaluated in the dynamically screened Born approximation.
In Fig.~\ref{fig:denseH-reinholz} we compare the practical formula of Reinholz et al. to our qLFP results for hydrogen at $\rho = \SI{1}{\gram\per\cubic\centi\meter}$.\footnote{%
  A lower density is used compared with our other results to prevent extrapolating the fit of Ref.~\cite{ReinholzPRE2015}.}
Reinholz et al.'s model predicts slightly weaker electron-electron scattering influence in the electrical conductivity at high temperature compared to our qLFP model, which is seen to approach the Spitzer-H\"arm value $R_{\sigma}\approx0.5816$ obtained from the classical Fokker-Planck equation~\cite{SpitzerPR1953}.
At the highest temperatures of the isochore shown, hydrogen is nondegenerate and weakly coupled, so the main physical difference between the approaches at high temperatures is dynamic screening, which is accounted for in Reinholz et al.'s calculations but not in qLFP~\cite{ReinholzPRE2015,DaligaultPOP2016}.
We also show qLFP results for the electron-electron scattering correction to the Lorentz gas thermal conductivity
\begin{equation}
  R_\lambda = \frac{\lambda}{\lambda(\text{no e-e})}
\end{equation}
and the corresponding Spitzer-H\"arm value $R_\lambda=0.2358$.}

%


\subsection{The Small-Angle Approximation}
\label{sec:small-angle}

\begin{figure*}[t]
  \centering
  \includegraphics[width=\textwidth]{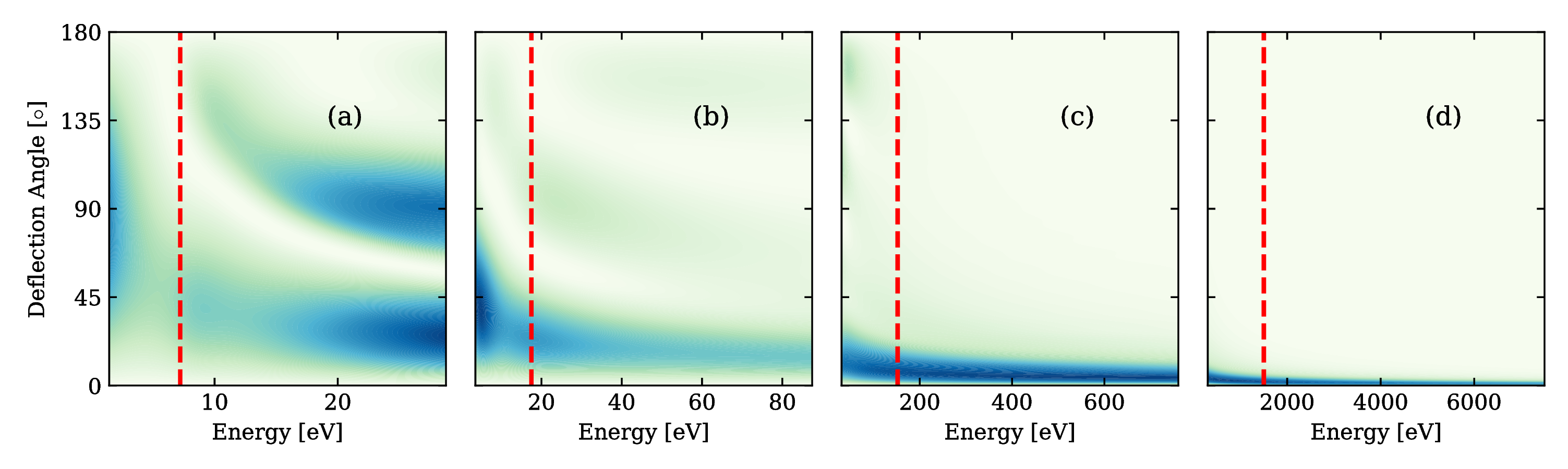}
  \includegraphics[width=\textwidth]{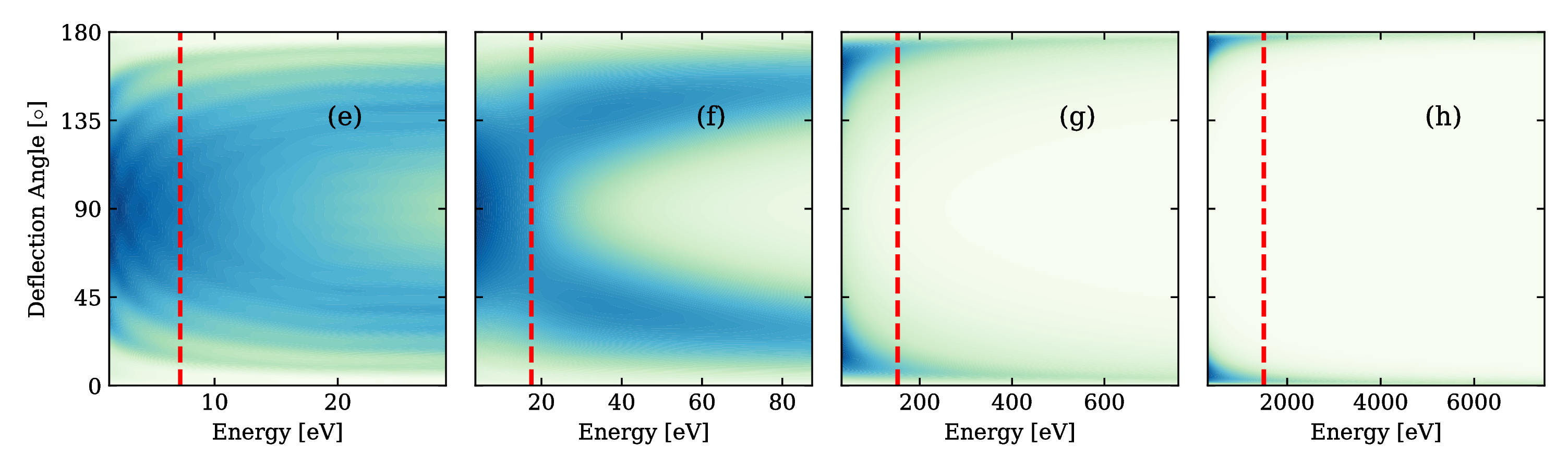}
  \caption{Distribution of deflection angles, $D_{ij}(E,\theta)$, for elastic electron-ion (upper) and electron-electron (lower) scattering in solid-density aluminum at temperatures of \SIlist[list-units=single]{1;10;100;1000}{\electronvolt} (left to right). The color scale in each panel shows only the relative magnitude of the angle distribution, with dark and light shades being large and small values of the distribution, respectively. The dashed red line marks the energy corresponding to the root-mean-square relative velocity at each temperature.}
  \label{fig:aluminum-dcs}
\end{figure*}
We now assess the small-angle approximation in greater detail by looking at the energy-resolved distribution of deflection angles for electron-ion and electron-electron mean-force scattering, plotted for solid-density aluminum in Fig.~\ref{fig:aluminum-dcs}.
The deflection angle distribution is related to the differential cross-sections by $D_{ij}(E,\theta) = 2\pi\sin\theta \dd{\sigma_{ij}}{\Omega}$, and its relative magnitude at large and small deflection angles gives an indication for whether scattering at a particular energy is mainly large- or small-angle.
The energies that contribute most to near-equilibrium transport are those corresponding to the root-mean-square relative momentum, which is indicated by the vertical dashed lines in each panel of Fig.~\ref{fig:aluminum-dcs}.
This value varies from $\frac35 \mu_e$ as $T\to0$ to $\frac32 k_BT$ as $T\to\infty$.
Near this energy, if $D_{ij}$ is peaked at small angles (less than \SI{45}{\degree}, say), then the qLFP theory is expected to be a good description of transport.
Note that the electron-electron cross-section is always symmetric about \SI{90}{\degree} due to the indistinguishability of electrons, so it is only necessary to consider the range $\theta<\SI{90}{\degree}$ in assessing the quality of the small-angle approximation for electron-electron scattering.

In the \SI{100}{\electronvolt} and \SI{1000}{\electronvolt} cases shown, both $D_{eI}$ and $D_{ee}$ are strongly forward-peaked, with almost all collisions involving deflections angles smaller than \SI{45}{\degree}.
Even as low as \SI{10}{\electronvolt}, electron-ion scattering involves mostly small deflections.
Large-angle electron-ion collisions at \SI{10}{\electronvolt} occur only at very low energies or due to resonance scattering, which appears as a \comment{faint band around $\SI{90}{\degree}$ in the figure.}
In contrast, electron-electron scattering at \SI{10}{\electronvolt} is predominantly large-angle.
The case of \SI{10}{\electronvolt} is especially important because Pauli blocking becomes important around this temperature.
It appears that at this temperature, qLFP accurately treats the Pauli blocking of electron-ion collisions but not electron-electron collisions (see \ref{sec:small-angle}).
At \SI{1}{\electronvolt}, both types of scattering are predominantly large-angle.
In addition, one sees prominent symmetry oscillations for electron-electron scattering due the interference between the forward- and backward-scattered electrons\cite{NewtonScattering}.

\section{Conclusions}
\label{sec:conc}

We have demonstrated that the qLFP collision theory, combined with accurate Coulomb logarithms based on mean force scattering, leads to predictions of electrical and thermal conductivity which are accurate over a wide range of temperatures relevant to dense plasmas.
Our calculations take electron-electron scattering into account on equal footing with electron-ion scattering, reproducing the classical result that electron-electron scattering is important to the conductivity of low-$Z$ materials.
We find that electron-electron scattering is important to the low-temperature behavior of the thermal conductivity as well, even at temperatures somewhat less than the Fermi energy.
It is only at very low temperatures that Pauli blocking eliminates the influence of electron-electron scattering on thermal conduction.
It is in this regime where the qLFP small-angle approximation finally breaks down in a way that cannot be recovered by our mean-force Coulomb logarithm model.

\begin{acknowledgments}
  We wish to thank J\'er\^ome Daligault for useful discussions.
  This work was performed under the auspices of the United States Department of Energy under Contract No.~89233218CNA000001.
\end{acknowledgments}

\appendix

\section{Relaxation-Time Approximation of the Uehling-Uhlenbeck Collision Integral}
\label{sec:trx-uu}

In Sec.~\ref{sec:coul-log}, it was argued that accurate electron-ion Coulomb logarithms could be inferred from the transport coefficients obtained from a relaxation-time collision operator.
In this Appendix, the relaxation-time collision operator is derived as an accurate approximation to the UU collision operator for the contribution of electron-ion scattering to near-equilibrium electron transport.

For electron-ion collisions, the UU collision integral is
\begin{equation}
  C_{eI}^{UU} = \int[f_e(\hat{\vec p}) \bar f_e(\vec p) f_I(\hat{\vec p}') - f_e(\vec p)\bar f_e(\hat{\vec p}) f_I(\vec p')] v_{Ie} d\sigma_{Ie}\frac{d\vec p'}{\omega_I}
\end{equation}
For the study of conduction by electrons, one can take the ion distribution to be an isotropic Maxwellian 
\begin{equation}
  f_I(\vec p') \approx f_I^{(0)}(p')
\end{equation}
and expand the electron distribution function in Legendre polynomials
\begin{equation}
  f_e(\vec p) \approx f_e^{(0)}(p) + \mu g(p)
\end{equation}
where $\mu$ is the cosine of the angle between $\vec p$ and the direction of transport, and $g(p)$ is the deviation from equilibrium.
For simplicity, it is assumed that the conduction force and temperature gradient are parallel so that the induced electric current and heat flux are also parallel.
The distribution function is assumed symmetric about this axis so that there is no azimuthal dependence.

Due to the smallness of $m_e/m_I$, electron-ion collisions involve negligible energy transfer between the electron and ion.
To a good approximation, one can say that $\hat p' = p'$ and $\hat p = p$, so that electron-ion collisions only rotate the electron's momentum vector.
Further, one may approximate the relative velocity by the electron's, $v_{Ie}\approx v = p/m_e$.
Retaining only terms up to $\bigO(\mu)$, noting that the isotropic term vanishes, and performing the integral over the ion momenta leaves
\begin{equation}
  C_{eI}^{UU} \approx n_Iv g(p)\int (\hat\mu - \mu) d\sigma_{Ie} 
\end{equation}
The change in $\mu$ due to a collision is given by $\hat\mu - \mu = -\mu(\cos\theta - 1)$, where $\theta$ is the deflection angle, which allows the remaining integral to be written in terms of the momentum-transfer cross-section
\begin{equation}
  \int (\hat\mu - \mu) d\sigma_{Ie} = -\mu\sigma_{eI}^{(1)}(v)
\end{equation}
Finally, since $\mu g = f_e - f_e^{(0)}$, we have reduced the UU collision operator to a relaxation-time approximation
\begin{equation}
  C_{eI}^{UU} \approx -\frac{f_e(\vec p) - f_e^{(0)}(p)}{\tau(p)}
\end{equation}
which is suitable for describing how collisions with ions affect electron transport.

\section{Corrections to Some Formulas Appearing in Ref.~\cite{DaligaultPOP2018}}
\label{app:typo}

In this Appendix, \comment{we} point out some minor errors in key formulas of Ref.~\cite{DaligaultPOP2018}.
Namely, Eqs.~(27a), (31), (32), (48), (49), (52), (56), and (59) contain unnecessary factors of $\beta \Pi/n$, which should instead be absorbed into the definition of the diffusion force, Eq.~(24).
Also, Eqs.~(73) and (74) have incorrect species labels on the chemical potentials, as can be seen by comparing with Daligault's Eqs.~(50) and (64) (which are correct) and noting how the partial bracket integrals assign species labels to their operands.
These errors do not affect Daligault's numerical results for the transport coefficients of plasmas in the nondegenerate limit, nor do they affect any of the results for the electron gas.
Our own calculations indicate that most of the qualitative results for degenerate plasmas still hold as well, they just cannot be used in quantitative comparisons with one's own implementation.

\section{Practical Distillation of the Chapman-Enskog Solution}
\label{app:ce}

A detailed derivation of the Chapman-Enskog solution to the qLFP equation and the corresponding expressions for transport coefficients may be found in Ref.~\cite{DaligaultPOP2018}.
In this Appendix, we summarize those aspects of the theory necessary for a practical implementation.

The Chapman-Enskog method seeks a solution to the qLFP kinetic equation near local thermal equilibrium.
The distribution function of each species is expanded to first order in an asymptotic series
\begin{equation}
  \label{eq:f-series}
  f_i = f_i^{(0)} + f_i^{(1)}
\end{equation}
where the deviation from local thermal equilibrium is written as
\begin{equation}
  \label{eq:f1}
  f_i^{(1)} = f_i^{(0)} [1 + \eta_i f_i^{(0)}] \phi_i
\end{equation}
with the unknown function $\phi_i$ having the form
\begin{equation}
  \label{eq:phi}
  \phi_i = -\frac1n \sum_{j=1}^K \vec D_i^j \cdot \vec d_j + \frac1n \vec A_i \cdot  \nabla\ln \beta
\end{equation}
where $\vec d_j$ is the diffusion force on species $j$ and $\vec D_i^j$ and $A_i$ are unknown functions of momentum, temperature, and chemical potential to be determined.
Once known, $\vec D_i^j$ and $\vec A_i$ determine the diffusive and thermal transport coefficients.
In general there is a third term for viscous transport, which we neglect.
Viscosity is dominated by the classical ions, for which accurate results based on the classical Boltzmann collision operator are already known\cite{DaligaultPRL2016}.

The coefficients $\vec A_i$ and $\vec D_i^j$ are the solutions to linear integral equations involving the linearized qLFP collision operator, $I_{ij}$,
\begin{subequations}
  \begin{equation}
    \sum_{j=1}^K \frac{n_in_j}{n} I_{ij}[\vec D^k] = \left(\delta_{ik} - \frac{\rho_i}{\rho}\right) \frac{\vec P}{m_i}f_i^{(0)} [1 + \eta_i f_i^{(0)}]
  \end{equation}
  \begin{equation}
    \sum_{j=1}^K \frac{n_in_j}{n} I_{ij}[\vec A] = \left[\frac{\beta P^2}{2m_i} - \frac{5 \qint_\frac32(\beta\mu_i)}{2 \qint_\frac12(\beta\mu_i)} \right] \frac{\vec P}{m_i} f_i^{(0)}[1 + \eta_i f_i^{(0)} ]
  \end{equation}
\end{subequations}
where $\vec P = \vec p - m_i \vec u$ is the momentum in the frame co-moving with the fluid at the local velocity $\vec u(\vec r, t)$.
The indices $i$ and $k$ run over all species labels.
The unknowns $\vec D^k$ and $\vec A$ are written without species subscripts because they are assigned by the operator $I_{ij}$, the detailed form of which not important for the present discussion but can be found in Ref.~\cite{DaligaultPOP2018}.
The solution to these integral equations is carried out by expanding $\vec A_i$ and $\vec D_i^j$ in a basis of orthogonal polynomials introduced by Daligault
\begin{equation}
  \label{eq:D-expansion}
  \vec D_i^j(\vec P) =  \frac12 \beta \vec P \sum_{p=0}^{r-1} d_{i,p}^{j,r} \qpoly_{\frac32,i}^{(p)}(\beta P^2/2m_i)
\end{equation}
\begin{equation}
  \label{eq:A-expansion}
  \vec A_i(\vec P) = -\frac12 \beta \vec P \sum_{p=0}^r a^r_{i,p} \qpoly_{\frac32,i}^{(p)}(\beta P^2/2m_i)
\end{equation}
The truncation of the expansions results in the so-called ``order-$r$'' Chapman-Enskog approximation for the transport coefficients.
The polynomials $\qpoly_{\nu,i}^{(n)}(x)$ are constructed to ensure optimal, monotonic convergence with respect to $r$.
They depend parametrically on the chemical potential
\begin{equation}
  \label{eq:qpoly}
  \qpoly_{\nu,i}^{(n)}(x) = \sum_{p=0}^n c_\nu^{n,p}(\beta\mu_i) x^p
\end{equation}
and the coefficients can be determined from the recurrence relation
\begin{equation}
  \label{eq:qpoly-coeffs}
  c_\nu^{n,p} = s_\nu^{n,p} - \sum_{q=p}^{n-1} c_\nu^{q,p} \frac{
    \sum_{i_n=0}^n\sum_{i_q=0}^q s_\nu^{n,i_n} c_\nu^{q,i_q} N_{\nu+i_n+i_q}
  }{
    \sum_{i_q=0}^q\sum_{j_q=0}^q c_\nu^{q,i_q} c_\nu^{q,j_q} N_{\nu+i_q+j_q}
  }
\end{equation}
initiated with $c_\nu^{0,0} = 1$.
The other quantities appearing in the recurrence relation are the Sonine polynomial coefficients
\begin{equation}
  \label{eq:sonine-coeff}
  s_\nu^{n,p} = \frac{(-1)^p}{ p!(n - p)! } \frac{\Gamma(\nu + n + 1)}{\Gamma(\nu + p + 1)}
\end{equation}
and numerical factor
\begin{equation}
  \label{eq:N-factor}
  N_\nu = \Gamma(\nu+1) \qint_{\nu-1}(\beta \mu)
\end{equation}
where the species index on the chemical potential is the same as for the polynomial being evaluated.

The expansions Eq.~\eqref{eq:D-expansion} and \eqref{eq:A-expansion} turn the integral equations for $\vec A_i$ and $\vec D_i^j$ into linear systems of equations for the coefficients $a^r_{i,p}$ and $d_{i,p}^{j,r}$.
Once obtained, these determine the mutual diffusivities
\begin{equation}
  \label{eq:Dij}
  [D_{ij}]_r = \frac1{2n} d_{i,0}^{j,r}
\end{equation}
thermal diffusivities
\begin{equation}
  \label{eq:Dti}
  [D_{Ti}]_r = -\frac1{2n} a^r_{i,0}
\end{equation}
and partial thermal conductivities
\begin{equation}
  \label{eq:lambda-primei}
  [\lambda'_i]_r = \frac{5k_B}{4} \left\{ \frac{7\qint_\frac52(\beta\mu_i)}{2\qint_\frac12(\beta\mu_i)} - \frac52\left[\frac{\qint_\frac32(\beta\mu_i)}{\qint_\frac12(\beta\mu_i)}\right]^{\!2}\right\} a^r_{i,1}
\end{equation}
where the square-bracket notation denotes the order-$r$ approximation to the transport coefficients.
The linear systems of equations to be solved for the expansion coefficients are
\begin{equation}
  \label{eq:d-system}
  \sum_{j=1}^K \sum_{q=0}^r L_{i,p}^{j,q} d_{j,q}^{k,r}   = \frac{8}{25k_B} \left(\delta_{ik} - \frac{\rho_i}{\rho} \right) \delta_{p0}
\end{equation}
\begin{equation}
  \label{eq:a-system}
  \sum_{j=1}^K \sum_{q=0}^r L_{i,p}^{j,q} a_{j,q}^r   = \frac{4}{5k_B} \left\{ \frac{7\qint_\frac52(\beta\mu_i)}{2\qint_\frac12(\beta\mu_i)} - \frac52\left[\frac{\qint_\frac32(\beta\mu_i)}{\qint_\frac12(\beta\mu_i)}\right]^{\!2}\right\} \delta_{p1}
\end{equation}
where the matrix elements are
\begin{equation}
  \label{eq:Lijpq}
  L_{i,p}^{j,q} = \Lambda_{i,p}^{j,q} - \frac{\rho_i}{\rho_K} \Lambda_{i,p}^{K,0} \delta_{q0} (1-\delta_{iK})
\end{equation}
\begin{widetext}
  \begin{equation}
    \label{eq:Lambdaijpq}
    \Lambda_{i,p}^{j,q}
    = \frac{8\sqrt{m_im_j}}{75k_B^2 T} \sum_{k_p=0}^p \sum_{k_q=0}^q 
    \left[
      \delta_{ij} c_\frac32^{p,k_p}(\beta\mu_i) c_\frac32^{q,k_q}(\beta\mu_i)  \sum_{h=1}^K \frac{n_i n_h}{n^2} A'_{ih,k_qk_p}
      \\
      + c_\frac32^{p,k_p}(\beta\mu_i) c_\frac32^{q,k_q}(\beta\mu_j) \frac{n_in_j}{n^2} A''_{ij,k_qk_p}
    \right] 
  \end{equation}
  \begin{equation}
    A'_{ij,pq} = \sum_{m=1}^{p+q} \left\{ [4pq + 2(p+q)] \binom{p+q-1}{m-1} + \binom{p+q}{m}  \right\}(p+q-m)!
    \mathcal A_{ij}^{m,p+q-m,0} + (p+q)! \mathcal A_{ij}^{0,p+q,0}
  \end{equation}
  \begin{multline}
    A''_{ij,pq} = -\sqrt{\frac{m_i}{m_j}} \Bigg( \sum_{m=1}^p \sum_{n=1}^q \left(\frac{m_j}{m_i}\right)^{\!m} \left\{ \binom{q-1}{n-1} \left[ 4pq \binom{p-1}{m-1} + 2q \binom{p}{m}\right]
    \right. \\ \left.
      + \binom{q}{n} \left[ 2p \binom{p-1}{m-1} + \binom{p}{m}\right]  \right\} (p-m)! (q-n)! \mathcal A_{ij}^{m+n, q-n, p-m}
    \\
    + \sum_{m=1}^p \left(\frac{m_j}{m_i}\right)^{\!m}\left[ 2p\binom{p-1}{m-1} + \binom{p}{m} \right](p-m)! q! \mathcal A_{ij}^{m,q,p-m}
    \\
    + \sum_{n=1}^q \left[2q \binom{q-1}{n-1} + \binom{q}{n}\right] p! (q-n)! \mathcal A_{ij}^{n,q-n,p}
    + p! q! \mathcal A_{ij}^{0,q,p} \Bigg)
  \end{multline}
  \begin{equation}
    \mathcal A_{ij}^{p,q,r} =  \Gamma_{ij} \frac{\beta^\frac32 m_j^\frac12}{2^\frac12 \pi m_i m_{ij}} \frac{1}{\qint_\frac12(\beta\mu_i) \qint_\frac12(\beta\mu_j)}
    \int_0^\infty x^{2p} \qint'_q(\beta\mu_i - x^2) \qint'_r(\beta\mu_j - \tfrac{m_j}{m_i}x^2) dx
  \end{equation}
\end{widetext}
where $Q'_\nu(z) = \dd{}{z}Q_\nu(z)$ is the derivative of the Fermi-Dirac integral.
These formulas are complicated but straightforward to implement.
We mention only a few technical points:
\begin{itemize}
\item One must use an efficient and reasonably robust implementation of the Fermi integrals, or equivalently polylogarithms. In our experience, Goano's algorithm (TOMS Algorithm 745) worked well\cite{GoanoTACM1995}.
  
\item The integrand for $\mathcal A_{ij}^{p,q,r}$ decays exponentially quickly at large $x$ and can be accurately truncated and evaluated with simple quadrature rules. The results shown in this work all truncated at $x=15$ and used a trapezoidal rule with a uniformly spaced mesh of 750 points, which was never appreciably different from truncating at $x=10$ and a mesh of 500 points.
  
\item We found it useful to exploit the symmetry relation $\mathcal A_{ij}^{p,q,r} = \left(\frac{m_j}{m_i}\right)^{1-p} \mathcal A_{ji}^{p,r,q}$ when $m_j > m_i$ to avoid large negative arguments of the Fermi function $\qint'_r(\beta\mu_j - \frac{m_j}{m_i}x^2)$, which we found was prone to floating-point underflow leading to spurious division by zero.
\end{itemize}

Eq.~\eqref{eq:d-system} and \eqref{eq:a-system} can be written in form that is more convenient for software implementation by introducing composite indices $(ip) = K i + p$ and $(jq) = K j + q$ to define a $Kr$-by-$Kr$ matrix $\mat L$ with elements
\begin{equation}
  L_{(ip)(jq)} = L_{i,p}^{j,q}
\end{equation}
and several length-$Kr$ vectors $\vec d^k$, $\vec a$, $\vec u^k$, and $\vec v$ with elements
\begin{equation}
  d^k_{(jq)} = d^{k,r}_{j,q}
\end{equation}
\begin{equation}
  a_{(jq)} = a_{j,q}^r
\end{equation}
\begin{equation}
  u^k_{(ip)} = \frac{8}{25k_B} \left( \delta_{ik} - \frac{\rho_i}{\rho} \right)\delta_{p0} 
\end{equation}
\begin{equation}
  v_{(ip)} =  \frac{4}{5k_B} \left\{ \frac{7\qint_\frac52(\beta\mu_i)}{2\qint_\frac12(\beta\mu_i)} - \frac52\left[\frac{\qint_\frac32(\beta\mu_i)}{\qint_\frac12(\beta\mu_i)}\right]^{\!2}\right\} \delta_{p1}
\end{equation}
Then the solution for the expansion coefficients is obtained by solving the $K+1$ independent linear systems
\begin{equation}
  \mat L \cdot \vec d^k = \vec u^k
\end{equation}
\begin{equation}
  \mat L \cdot \vec a = \vec v
\end{equation}
The transport coefficients may then be immediately evaluated from Eq.~\eqref{eq:Dij}, \eqref{eq:Dti}, and \eqref{eq:lambda-primei}.

The more familiar electrical and thermal conductivities may be obtained from these.
The electrical conductivity is
\begin{equation}
  \label{eq:sigma}
  [\sigma]_r = \frac{1}{nk_BT}\sum_{i=1}^K \sum_{j=1}^K e_in_i e_jn_j [D_{ij}]_r
\end{equation}
the thermal conductivity is
\begin{equation}
  \label{eq:lambda}
  [\lambda]_r = \sum_{i=1}^K \left( \frac{n_i}{n} [\lambda'_i]_r - nk_B [k_{Ti}]_r [D_{Ti}]_r \right)
\end{equation}
and the electronic contribution to the thermoelectric power is 
\begin{equation}
  [\alpha]_r = -\frac{k_B}{e} \left( \frac{n}{n_e} [k_{Te}]_r + \frac{5\qint_\frac32(\beta\mu_e)}{2\qint_\frac12(\beta\mu_e)}\right)
\end{equation}
The thermal conductivity and thermoelectric power require the thermodiffusion ratios which are determined by solving
\begin{equation}
  \sum_{j=1} [D_{ij}]_{r+1} [k_{Tj}]_r = [D_{Ti}]_r
\end{equation}
with the constraint $\sum_j [k_{Tj}]_r = 0$.
Note that an order-$r$ approximation to the thermal conductivity and thermoelectric power requires an order-$(r+1)$ approximation of the electrical conductivity.
This is because the Chapman-Enskog ``order'' regrettably refers to the number of polynomials retained in Eq.~\eqref{eq:D-expansion} and Eq.~\eqref{eq:A-expansion}, rather than the highest degree of polynomial.
The offset by one order just reflects that a consistent calculation of thermal and electrical conductivities should use the same truncated polynomial basis for each.

Finally, we list several useful symmetry properties and constraints that are useful in checking a software implementation of the Chapman-Enskog solution.
\begin{itemize}
\item Species-interchange symmetry of the $\Lambda$ matrix elements:
  $\Lambda_{i,p}^{j,q} = \Lambda_{j,q}^{i,p}$
  
\item Symmetry of the mutual diffusion coefficient matrix:
  $D_{ij} = D_{ji}$
  
\item Positivity of the diagonal elements of the mutual diffusion coefficient matrix:
  $ D_{ii} > 0 $

\item Momentum conservation constraints on the mutual and thermal diffusion coefficients:
  $ \sum_{j=1}^K \frac{\rho_j}{\rho} D_{ij} = 0 $ and 
  $ \sum_{i=1}^K \frac{\rho_i}{\rho} D_{Ti} = 0 $

\end{itemize}
%

\bibliography{refs}

\end{document}